\documentclass[modern]{aastex61}

\usepackage[colorinlistoftodos]{todonotes} 

\usepackage{cancel}		

\newcommand{\be}{\begin{eqnarray}}
\newcommand{\ee}{\end{eqnarray}}

\newcommand{\frb}{FRB~121102}

\newcommand{\RM}{{\rm RM}}

\newcommand{\RMunits}{{\rm rad~m^{-2}}}

\newcommand{\DM}{{\rm DM}}

\newcommand{\DMhost}{\DM_{\rm host}}
\newcommand{\DMRM}{{\DM_{\RM}}}
\newcommand{\DMunits}{{\rm pc~cm^{-3}}}

\newcommand{\aau}{a_{\rm AU}}

\newcommand{\dsl}{d_{\rm sl}}

\newcommand{\dso}{d_{\rm so}}

\newcommand{\lau}{l_{\rm AU}}
\newcommand{\nul}{\nu_{\,l}}

\received{}
\revised{}
\accepted{}
\submitjournal{ApJL}

\shorttitle{FRB~121102 Burst Structure}
\shortauthors{Hessels et al.}

\begin{document}

\title{FRB~121102 Bursts Show Complex Time-Frequency Structure}

\correspondingauthor{J.~W.~T.~Hessels}
\email{J.W.T.Hessels@uva.nl}

\author{J.~W.~T.~Hessels}
\affiliation{ASTRON, Netherlands Institute for Radio Astronomy, Oude Hoogeveensedijk 4, 7991~PD Dwingeloo, The Netherlands}
\affiliation{Anton Pannekoek Institute for Astronomy, University of Amsterdam, Science Park 904, 1098~XH Amsterdam, The Netherlands}
\author{L.~G.~Spitler}
\affiliation{Max-Planck-Institut f\"ur Radioastronomie, Auf dem H\"{u}gel 69, D-53121 Bonn, Germany}
\author{A.~D.~Seymour}
\affiliation{Green Bank Observatory, PO Box 2, Green Bank, WV 24944, USA}
%
\author{J.~M.~Cordes}
\affiliation{Cornell Center for Astrophysics and Planetary Science and Department of Astronomy, Cornell University, Ithaca, NY 14853, USA}
\author{D.~Michilli}
\affiliation{ASTRON, Netherlands Institute for Radio Astronomy, Oude Hoogeveensedijk 4, 7991~PD Dwingeloo, The Netherlands}
\affiliation{Anton Pannekoek Institute for Astronomy, University of Amsterdam, Science Park 904, 1098~XH Amsterdam, The Netherlands}
\author{R.~S.~Lynch}
\affiliation{Green Bank Observatory, PO Box 2, Green Bank, WV 24944, USA}
\affiliation{Center for Gravitational Waves and Cosmology, West Virginia University, Morgantown, WV 26506}
\author{K.~Gourdji}
\affiliation{Anton Pannekoek Institute for Astronomy, University of Amsterdam, Science Park 904, 1098~XH Amsterdam, The Netherlands}
\author{A.~M.~Archibald}
\affiliation{Anton Pannekoek Institute for Astronomy, University of Amsterdam, Science Park 904, 1098~XH Amsterdam, The Netherlands}
\affiliation{ASTRON, Netherlands Institute for Radio Astronomy, Oude Hoogeveensedijk 4, 7991~PD Dwingeloo, The Netherlands}
\author{C.~G.~Bassa}
\affiliation{ASTRON, Netherlands Institute for Radio Astronomy, Oude Hoogeveensedijk 4, 7991~PD Dwingeloo, The Netherlands}
\author{G.~C.~Bower}
\affiliation{Academia Sinica Institute of Astronomy and Astrophysics, 645 N. A'ohoku Place, Hilo, HI 96720, USA}
\author{S.~Chatterjee}
\affiliation{Cornell Center for Astrophysics and Planetary Science and Department of Astronomy, Cornell University, Ithaca, NY 14853, USA}
\author{L.~Connor}
\affiliation{ASTRON, Netherlands Institute for Radio Astronomy, Oude Hoogeveensedijk 4, 7991~PD Dwingeloo, The Netherlands}
\affiliation{Anton Pannekoek Institute for Astronomy, University of Amsterdam, Science Park 904, 1098~XH Amsterdam, The Netherlands}
\author{F.~Crawford}
\affiliation{Dept. of Physics and Astronomy, Franklin and Marshall College, Lancaster, PA 17604-3003, USA}
\author{J.~S.~Deneva}
\affiliation{George Mason University, resident at the Naval Research Laboratory, Washington, DC 20375, USA}
\author{V.~Gajjar}
\affiliation{Space Sciences Laboratory, 7 Gauss way, University of California, Berkeley, CA 94720, USA}
\author{V.~M.~Kaspi}
\affiliation{Department of Physics and McGill Space Institute, McGill University, Montreal, QC, Canada H3A 2T8}
\author{A.~Keimpema}
\affiliation{Joint Institute for VLBI ERIC, Oude Hoogeveensedijk 4, 7991~PD Dwingeloo, The Netherlands}
\author{C.~J.~Law}
\affiliation{Department of Astronomy and Radio Astronomy Lab, University of California, Berkeley, CA 94720, USA}
\author{B.~Marcote}
\affiliation{Joint Institute for VLBI ERIC, Oude Hoogeveensedijk 4, 7991~PD Dwingeloo, The Netherlands}
\author{M.~A.~McLaughlin}
\affiliation{Department of Physics and Astronomy, West Virginia University, Morgantown, WV, 26501}
\affiliation{Center for Gravitational Waves and Cosmology, West Virginia University, Morgantown, WV 26506}
\author{Z.~Paragi}
\affiliation{Joint Institute for VLBI ERIC, Oude Hoogeveensedijk 4, 7991~PD Dwingeloo, The Netherlands}
\author{E.~Petroff}
\affiliation{ASTRON, Netherlands Institute for Radio Astronomy, Oude Hoogeveensedijk 4, 7991~PD Dwingeloo, The Netherlands}
\affiliation{Anton Pannekoek Institute for Astronomy, University of Amsterdam, Science Park 904, 1098~XH Amsterdam, The Netherlands}
\author{S.~M.~Ransom}
\affiliation{National Radio Astronomy Observatory, Charlottesville, VA 22903, USA}
\author{P.~Scholz}
\affiliation{Dominion Radio Astrophysical Observatory, Herzberg Astronomy \& Astrophysics Research Centre, National Research Council Canada, P.O. Box 248, Penticton, V2A 6J9, Canada}
\author{B.~W.~Stappers}
\affiliation{Jodrell Bank Center for Astrophysics, School of Physics and Astronomy, University of Manchester, Oxford Road Manchester, UK}
\author{S.~P.~Tendulkar}
\affiliation{Department of Physics and McGill Space Institute, McGill University, Montreal, QC, Canada H3A 2T8}



\begin{abstract}
\frb\ is the only known repeating fast radio burst source.  Here we analyze a wide-frequency-range ($1-8$\,GHz) sample of high-signal-to-noise, coherently dedispersed bursts detected using the Arecibo and Green Bank telescopes.  These bursts reveal complex time-frequency structures that include sub-bursts with finite bandwidths.  The frequency-dependent burst structure complicates the determination of a dispersion measure (DM); we argue that it is appropriate to use a DM metric that maximizes frequency-averaged pulse structure, as opposed to peak signal-to-noise, and find ${\rm DM} = 560.57 \pm 0.07$\,pc\,cm$^{-3}$ at MJD~57644.  After correcting for dispersive delay, we find that the sub-bursts have characteristic frequencies that typically drift lower at later times in the total burst envelope.  In the $1.1-1.7$\,GHz band, the $\sim 0.5-1$-ms sub-bursts have typical bandwidths ranging from $100-400$\,MHz, and a characteristic drift rate of $\sim 200$\,MHz/ms towards lower frequencies.  At higher radio frequencies, the sub-burst bandwidths and drift rate are larger, on average.  While these features could be intrinsic to the burst emission mechanism, they could also be imparted by propagation effects in the medium local to the source.  Comparison of the burst DMs with previous values in the literature suggests an increase of $\Delta{\rm DM} \sim 1-3$\,pc\,cm$^{-3}$ in 4 years, though this could be a stochastic variation as opposed to a secular trend.  This implies changes in the local medium or an additional source of frequency-dependent delay.  Overall, the results are consistent with previously proposed scenarios in which \frb\ is embedded in a dense nebula.
\end{abstract}




\section{Introduction} \label{sec:intro}

Fast radio bursts (FRBs) are short-duration astronomical radio flashes of apparent extragalactic origin \citep{lbm+07,tsb+13,pbj+16}.  FRB emission arrives later at lower radio frequencies, and this has been attributed to dispersive delay from intervening ionised material.  This dispersive delay is quadratic with radio frequency ($\Delta t \propto \nu^{-2}$), and its magnitude is proportional to the dispersion measure (DM), which is the column density of free electrons between source and observer.  The large DMs of FRBs are inconsistent with models of the Galactic free electron density distribution \citep{cl02,ymw17}.  This suggests that FRBs originate at extragalactic distances, because their anomalously large DMs can not be explained by an additional dispersive delay from material local to a source in the Milky Way but can be explained by material in a host galaxy and the intergalactic medium \citep{lbm+07,tsb+13}.

Discovered in the Arecibo PALFA pulsar survey \citep{cfl+06,lbh+15}, \frb\ is a source of sporadically repeating fast radio bursts \citep{sch+14,ssh+16a,ssh+16b}.  The direct and precise localization of these bursts has shown that \frb\ is hosted in the star-forming region of a dwarf galaxy at a luminosity distance of $\sim 1$\,Gpc \citep[$z = 0.193$;][]{clw+17,tbc+17,mph+17,bta+17}.  This association thus confirms the extragalactic distance of \frb, as was previously inferred from its DM \citep{sch+14}.  \frb\ is also associated with a compact (diameter $< 0.7$\,pc), persistent radio source with isotropic luminosity $L_{\rm radio} \sim 10^{39}$\,erg\,s$^{-1}$ \citep{clw+17,mph+17}.
Deep X-ray and $\gamma$-ray observations have found no persistent high-energy counterpart to \frb\ \citep{sbh+17}. 
Many models for \frb\ have focused on a young, energetic and highly magnetized neutron star origin \citep[e.g.][]{csp16,cw16,lbp16}.  Based on the host galaxy type, which is also known to host superluminous supernovae (SLSNe) and long gamma-ray bursts (LGRBs), it has been suggested that \frb\ originates from a millisecond magnetar formed in the last few decades \citep{mbm17,tbc+17,mph+17}.  This scenario can also naturally explain the co-location of \frb\ with a star-forming region, as well as its association with the persistent radio source, which would represent a pulsar or magnetar wind nebula (PWN or MWN) and/or a supernova remnant (SNR) \citep{pir16,mkm16,km17,mmb+18}.


As yet, no other FRB source has been seen to repeat, despite dedicated searches for additional bursts \citep[e.g.,][]{pjk+15,rsj15,smb+18}, nor are there any other definitive host galaxy associations.  While \citet{kjb+16} present a potential afterglow to FRB~150418, \citet{wb16} argue that the putative counterpart is unassociated variability of an active galactic nucleus in the same field \citep[see also discussion in][]{bbt+16,jkb+17}.  Thus, it remains unclear whether \frb\ has a similar physical origin to other known FRBs \citep[e.g.,][]{rav18}.


Optical, X-ray and $\gamma$-ray observations that are simultaneous with detected \frb\ radio bursts have failed to identify any prompt high-energy counterpart to the radio bursts themselves \citep{dfm+16,hds+17,sbh+17}.  Given the absence of multi-wavelength counterparts, the properties of the radio bursts are thus critical for understanding the emission mechanism \citep{bel17,lyub14,lyut17,wax17} and the local environment of the source through imparted propagation effects \citep{cwh+17}.  The bursts have typical durations of milliseconds, but also show fine structure as narrow as $\sim 30$\,$\mu$s \citep{msh+18}.  The spectrum varies between bursts, even those that are separated by minutes or less \citep[e.g., Fig.~3 of][]{gsp+18}.  Simultaneous, multi-telescope data show that some bursts are visible over a relatively broad range of frequencies \citep[$> 1$\,GHz, see][]{lab+17}.  However, wide-band observations also show that many of the bursts peak in brightness within the observing band and are not well modeled by a power law \citep{ssh+16a,ssh+16b}.  

  
Recently, the detection of \frb\ bursts at relatively high radio frequencies of $4-8$\,GHz has revealed that the bursts are $\sim 100$\% linearly polarized, with a flat polarization position angle across the bursts; no circular polarization is detected \citep{msh+18,gsp+18}.  This provides new clues about the emission mechanism, and allows a more detailed phenomenological comparison to be made with other known types of millisecond-duration astronomical radio signals --- including various forms of pulsar and magnetar pulsed radio emission, which are often highly polarized \citep[e.g.,][]{gl98,efk+13}.  The polarized signal also reveals that an extreme Faraday rotation is imparted on the bursts: the rotation measure (RM) in the source frame was RM$_{\rm src} = 1.46 \times 10^5$\,rad\,m$^{-2}$ at the first epoch of detection, and was 10\% lower 7 months later \citep{msh+18,gsp+18}. This shows that \frb\ is in an extreme and dynamic magneto-ionic environment --- e.g., the vicinity of an accreting massive black hole (MBH) or within a highly magnetized PWN/MWN and SNR. The properties of the aforementioned persistent radio source are consistent with both these scenarios, as are the constraints from the non-detections of persistent high-energy emission \citep{clw+17,tbc+17,mph+17,sbh+17}.  


Here we present a multi-frequency subset of high-signal-to-noise \frb\ bursts that better demonstrate the complex time-frequency structure hinted at by previously reported bursts in the literature \citep[e.g.,][]{ssh+16a,ssh+16b,sbh+17}. These add substantial observational clues for modeling the underlying emission mechanism and propagation effects imparted near the source.  In \S\ref{sec:obs} we present the observations and selection of the burst sample.  We analyse the time-frequency properties of this sample in \S\ref{sec:analysis}, and discuss possible consequences for understanding \frb, and the FRBs in general, in \S\ref{sec:disc}.  Lastly, in \S\ref{sec:conc} we conclude and provide an outlook to future work inspired by the results presented here.

\section{Observations and Burst Sample} \label{sec:obs}

\subsection{Arecibo and GBT Observational Configurations}


Until recently, the available time and frequency resolution of FRB detections has been a limitation in studying their properties.  Even in the case of real-time detections, dedispersion of the bursts has typically been done incoherently \citep[though see][]{ffb+18}, meaning that there is residual time smearing from intra-channel delays \citep{pbb+15,kjb+16}.  The known DM of \frb\ allows for coherent de-dispersion\footnote{A method that completely corrects for intra-channel smearing from dispersive delay; see \citet{hr75} and \citet{lk04}.}, and the precise localization allows observations up to much higher frequencies (where the telescope field-of-view is narrower) compared to all other known FRB sources \citep{gsp+18}.


Arecibo observations (project P3094) were performed with the L-Wide receiver, which provides a $1150-1730$\,MHz band, dual linear polarizations, a gain $G \sim 10.5$\,K/Jy, and a receiver temperature $T_{\rm sys} \sim 30$\,K.  Coherently dedispersed filterbank data with full Stokes information were recorded using the PUPPI backend \citep[a clone of the GUPPI backend, described in][]{drd+08}.  Before each integration on \frb, we also acquired a 60-s calibration scan for polarimetric calibration.  The 8-bit data provide 10.24-$\mu$s time resolution and 1.5625-MHz spectral channels.  These channels were coherently dedispersed online to a fiducial DM$_{\rm fid} = 557.0$\,pc~cm$^{-3}$.  Hence, any residual intra-channel dispersive smearing is negligible as long as this is close to the true DM of the bursts: for deviations, $\Delta$DM$_{\rm fid}$, from DM$_{\rm fid}$ the residual temporal smearing scales as $\sim 4 \times \Delta$DM$_{\rm fid}$\,$\mu$s --- i.e., DM smearing is $\lesssim 20$\,$\mu$s in these data.
For comparison, the intra-channel DM smearing in the original \frb\ burst detections made with the Arecibo Mock Spectrometers was $700$\,$\mu$s \citep{sch+14,ssh+16a}.


Green Bank Telescope (GBT) observations (projects GBT16B-391, GBT17A-319) used the S-band receiver, with a $1600-2400$\,MHz band, dual linear polarizations, a gain $G \sim 2$\,K/Jy, and a receiver temperature $T_{\rm sys} \sim 25$\,K.  Data were recorded with the GUPPI backend \citep{drd+08} in an identical observing mode, and with the same time/frequency resolutions and polarimetric calibration scans as those described above for Arecibo/PUPPI. 

\subsection{Selection of Burst Sample}


To search the Arecibo coherently dedispersed filterbank data for bursts, we first used {\tt psrfits\_subband} from psrfits\_utils\footnote{https://github.com/demorest/psrfits\_utils} to subband and downsample the raw data to 12.5\,MHz frequency channels and 81.92\,$\mu$s total intensity (Stokes I) time samples.  Using the PRESTO\footnote{https://github.com/scottransom/presto} \citep{ran01} tool {\tt prepsubband}, we then created dedispersed time series (summed over the full 800-MHz frequency band), using a range of trial DMs from $461-661$\,pc\,cm$^{-3}$, in steps of $1$\,pc\,cm$^{-3}$. 
The GBT data were processed in a very similar way, but in this case the subbanded data used 40.96\,$\mu$s time samples and kept the full 1.56-MHz frequency resolution, while the DM trials were for a range of $527-587$\,pc\,cm$^{-3}$ and step size $0.1$\,pc\,cm$^{-3}$.

In both cases, the dedispersed timeseries were searched for single pulses above a 6-$\sigma$ threshold using PRESTO's {\tt single\_pulse\_search.py}.
We chose not to apply a radio frequency interference (RFI) mask in this process in order to avoid the possibility of rejecting a very bright and relatively narrow-band burst.
The dynamic spectra (radio frequency versus time) of candidate single-pulse events were inspected by eye to differentiate genuine astrophysical bursts from RFI.


The 1.4-GHz Arecibo sample presented here was detected during a high-cadence observing campaign in 2016 September \citep{clw+17,lab+17}.  Specifically, the sample was selected by choosing bursts with S/N $> 60$, as reported by {\tt single\_pulse\_search.py}, which searches a range of pulse widths using a boxcar matched filter.  This S/N is averaged over the full band and corresponds to an equivalent fluence limit of $> 0.2$\,Jy\,ms, assuming a 1-ms-wide burst.   
The S/N threshold was chosen in order to select just the brightest detected bursts, but to also retain a sufficiently large sample for comparison.  A complementary sample of Arecibo bursts observed at 4.5\,GHz, using the identical PUPPI recording setup, is presented in \citet{msh+18}.  We do not include a re-analysis of those bursts here because the available fractional observing bandwidth ($\sim 15$\%) is significantly lower compared to the data presented here, and insufficient to accurately study their broadband spectral behavior (see also discussion below).


The 2.0-GHz GBT bursts are from 2016 September and 2017 July and were also selected to have S/N $> 60$ (this corresponds to an equivalent fluence limit of $> 0.8$\,Jy\,ms, assuming a 1-ms-wide burst).  We chose an identical S/N threshold as for the Arecibo selection, in order to have comparable sensitivity to faint structures in the bursts.  To complement the Arecibo and GBT bursts, we also include in the sample a highly structured burst observed over an ultra-wide band of $4.6-8.2$\,GHz with the GBT as part of the Breakthrough Listen (BL) project\footnote{These data are available to download at http://seti.berkeley.edu/frb121102/.} \citep[for further details of the observational setup and analysis used to detect that burst, see][]{gsp+18}.


The full sample considered here is summarized in Table~\ref{tab:bursts} along with,  as a point of comparison,  the earliest 1.4-GHz \frb\ burst detected using coherent dedispersion \citep{ssh+16b}.  For each of the selected bursts, we used {\tt dspsr} \citep{Str11} to extract a window of full-resolution, full-Stokes raw data around the nominal burst time\footnote{These data products are available via Dataverse.} and produced a dedispersed dynamic spectrum using tools from PSRCHIVE\footnote{http://psrchive.sourceforge.net/} \citep{Str12}.  We then manually excised narrow-band RFI (channels with excess power before and/or after the burst), blanked recorded channels beyond the edges of the receiver band, and applied a bandpass correction using tools from PSRCHIVE.  The resulting dynamic spectra of the bursts are shown in Figure~\ref{fig:weird_bursts}\footnote{Three-dimensional printable models of these data cubes are freely available at https://www.thingiverse.com/thing:2723399.}.  They reveal a variety of temporal and spectral features, and in the rest of the paper we will refer to bright, relatively isolated patches in time-frequency as `sub-bursts'.  Note that the narrow-band, horizontal stripes in these dynamic spectra are due predominantly to RFI excision, which is necessary in order to reveal faint features in the bursts (the exception is GB-BL, where scintillation is also visible).  We analyze the time-frequency properties of the bursts and their sub-bursts in \S\ref{sec:analysis}.

We note that selecting only bursts with large S/N possibly introduces a bias towards more complex structure, if this structure is typically faint compared to the brightest peak in a burst.  That may contribute to why the bursts in the sample presented here are typically more complex in morphology compared to the entire sample of bursts detected and reported so far \citep[e.g,][]{ssh+16a,ssh+16b}.  However, we also note that high-S/N, relatively unstructured bursts have been detected from \frb\ \citep[e.g.,][]{sbh+17,mph+17}, and the sub-bursts within a burst are often of comparable brightness.  This suggests that any such bias is not strong.

\section{Analysis \& Results} \label{sec:analysis}

Here we present the properties of the burst sample defined in \S\ref{sec:obs}.

\subsection{DM Ambiguities}
\label{sec:dm_amb}

The dispersive delays across the Arecibo 1.4-GHz and GBT 2.0-GHz bands are roughly $1.0$\,s and $0.5$\,s, respectively, for ${\rm DM}_{\rm fid} = 557$\,pc\,cm$^{-3}$.  The dynamic spectra shown in Figure~\ref{fig:weird_bursts} are corrected using our best estimate of the dispersive delay.  However, there is an ambiguity between burst structure and DM because of the evolving burst morphology with radio frequency.  For example, a frequency-dependent profile shift on the order of 1\,ms can influence the measured DM at the 0.5\,pc\,cm$^{-3}$ level, and this is easily detectable, even by eye.  
Furthermore, intrinsically frequency-dependent emission time or local propagation effects can also possibly influence the apparent DM.  Hence, while a large fraction ($> 99$\%) of the frequency-dependent arrival time delay is likely due to dispersion in the intervening Galactic, intergalactic and host galaxy medium, there may also be additional non-dispersive effects that are difficult to distinguish from DM.

Before we can analyze the time-frequency properties of the bursts in detail, we must decide on an appropriate metric for determining DM.  We argue that choosing a DM that maximizes the peak S/N of the bursts is incorrect in this case.  Instead, we search a range of trial DMs and, effectively, we determine at what DM value each sub-burst is correctly de-dispersed individually, i.e. all significant emission in the sub-burst arrives simultaneously.  This makes the basic assumption that burst temporal components each emit simultaneously over a broad range of frequencies; a different underlying assumption, e.g. that there is an intrinsic, frequency-dependent delay in emission time, could also be considered.  Furthermore, here we determine a single DM per burst, and do not attempt to apply this metric to individual sub-bursts, which could have different apparent DMs in certain scenarios, as we discuss below.

To find an optimal DM under these assumptions, we maximize the steepness, i.e. time derivative, of peaks in the frequency-averaged burst profile.  Specifically, we search for the DM that maximizes the mean square of each profile's forward difference time derivative\footnote{For a similar approach, see \citet{gsp+18}.}. Because these time derivatives are susceptible to noise, and since we are searching for features that vary with DM, a two-dimensional Gaussian convolution (with $\sigma_{\rm DM} = 0.08$\,pc\,cm$^{-3}$ and $\sigma_{\rm time} = 82$\,$\mu$s) is performed within the DM versus time space before squaring and averaging over the time axis.  The resulting mean squared versus DM curve is then fitted with a high-order polynomial, and the peak DM value is then interpolated from this fit (Figure~\ref{fig:burst_DM}).

This is roughly the same as maximizing the structure in the frequency-averaged burst profile.  We find that all the 1.4-GHz and 2.0-GHz bursts in this sample are well modeled by a ${\rm DM} \sim 560.5$\,pc~cm$^{-3}$ (Table~\ref{tab:bursts}).  In contrast, maximizing the peak S/N of each burst leads to sub-bursts that overlap in time and sweep upward in frequency, as well as displaying a broader range of apparent DMs \citep[see also Fig.~1 of][]{gsp+18}.  The AO-01 to AO-13 bursts span a time range of only 11 days, and for 8 of these it was possible to derive a structure-maximizing DM (for the others, the method did not converge).  The average DM of these bursts is 560.57\,pc\,cm$^{-3}$, with a standard deviation of 0.07\,pc\,cm$^{-3}$ --- only marginally larger than the formal uncertainties on the individual DM determinations.  Given how well a single DM per burst aligns the sub-bursts such that they each arrive at the same time across the frequency band, we estimate that variations in apparent DM between sub-bursts are $\lesssim 0.1$\,pc\,cm$^{-3}$.  In contrast, for these same 8 bursts, the DMs from maximizing peak S/N are systematically higher, with an average of 562.58\,pc\,cm$^{-3}$ and a much larger standard deviation of 1.4\,pc\,cm$^{-3}$.  The much smaller scatter in DMs from the structure-maximizing metric arguably further justifies that approach; however, given the extreme magneto-ionic environment of the source \citep{msh+18}, we cannot rule out that there are relatively large DM variations between bursts.

\subsection{DM Variability}

The complex and frequency-dependent burst profiles show that adequate time resolution is critical in determining accurate DMs for \frb\ and, by extension, whether DM varies with epoch.  A ${\rm DM} = 560.57 \pm 0.07$\,pc\,cm$^{-3}$ at MJD~57644 (the average epoch of bursts AO-01 to AO-13) is roughly compatible with the range ${\rm DM} = 558.1 \pm 3.3$\,pc\,cm$^{-3}$ found by \citet{ssh+16a} --- i.e. the earliest sample of detected bursts from MJD~57159 and MJD~57175.  However, those data were only incoherently dedispersed, and hence unresolved burst structure may be the cause of the apparent spread in DMs in the \citet{ssh+16a} sample.  Furthermore, those DMs were determined using a S/N-maximizing metric, and hence are overestimated if there was unresolved, frequency-dependent sub-burst structure like that seen in the sample presented here.  

In the upper-left panel of Figure~\ref{fig:weird_bursts}, we show the dynamic spectrum of AO-00, the earliest 1.4-GHz burst from \frb\ detected using coherent dedispersion \citep[first presented as `burst 17' in][]{ssh+16b}, as it appears dedispersed to 560.5\,pc\,cm$^{-3}$.
The optimal DM value for MJD~57644 appears to be marginally too high for this burst from MJD~57364, where \citet{ssh+16b} found the optimal value to be 558.6\,$\pm 0.3 \pm 1.4$ \,pc\,cm$^{-3}$.  This value optimizes both peak S/N and burst structure; here the quoted uncertainties are, in order, statistical and systematic, where the systematic uncertainty was based on measuring the $\Delta$DM that results in a DM delay across the band equal to half the burst width.  However, because this burst was coherently dedispersed, we argue that it is unnecessary to consider this additional systematic uncertainty, which was added to account for possible frequency-dependent profile evolution.  In summary, comparing the burst DMs in the sample here with those of the earliest detections suggests that the DM of \frb\ has increased by $\sim 1-3$\,pc\,cm$^{-3}$ ($\sim 0.2-0.5$\% fractional) in the 4 years since its discovery, but we caution that there could be stochastic variations on shorter timescales and that this is not necessarily a secular trend.

\subsection{Polarimetry}

The recent detection of \frb\ bursts at relatively high radio frequencies \citep[$4-8$\,GHz;][]{gsm+17,msh+18,gsp+18,shb+18} has enabled the detection of a high linear polarization fraction (L/I $\sim 100$\%), no detectable circular polarization ($|\rm{V}|$/I $ \sim 0$\%), and an exceptionally large Faraday rotation measure (RM$_{\rm src} = 1.46 \times 10^5$\,rad\,m$^{-2}$).  Bandwidth smearing (intra-channel phase wrapping) in the 1.5-MHz channels at frequencies $< 2.4$\,GHz explains why previous polarization searches have been unsuccessful, if the observer frame RM was $\gtrsim 10^5$\,rad\,m$^{-2}$ at those epochs.  Additionally, it is possible that \frb\ is less polarized at lower frequencies.  For the 1.4-GHz and 2.0-GHz bursts presented here, we nonetheless searched for polarized emission using PSRCHIVE's {\tt rmfit} routine to investigate a range $|{\rm RM}| < 3 \times 10^5$\,rad\,m$^{-2}$ after a basic polarimetric calibration \citep[see][for details]{msh+18}.  This was to check whether the RM was perhaps much lower at earlier epochs, but again no linearly or circular polarized emission was detected above a 3-$\sigma$ significance.  The polarimetric properties of the high-frequency burst GB-BL (Table~\ref{tab:bursts}, Figure~\ref{fig:weird_bursts}) are presented in \citet{gsp+18}.

\subsection{Time-Frequency Burst Analysis}

\begin{figure*}
\centerline{\includegraphics[width=0.95\textwidth]{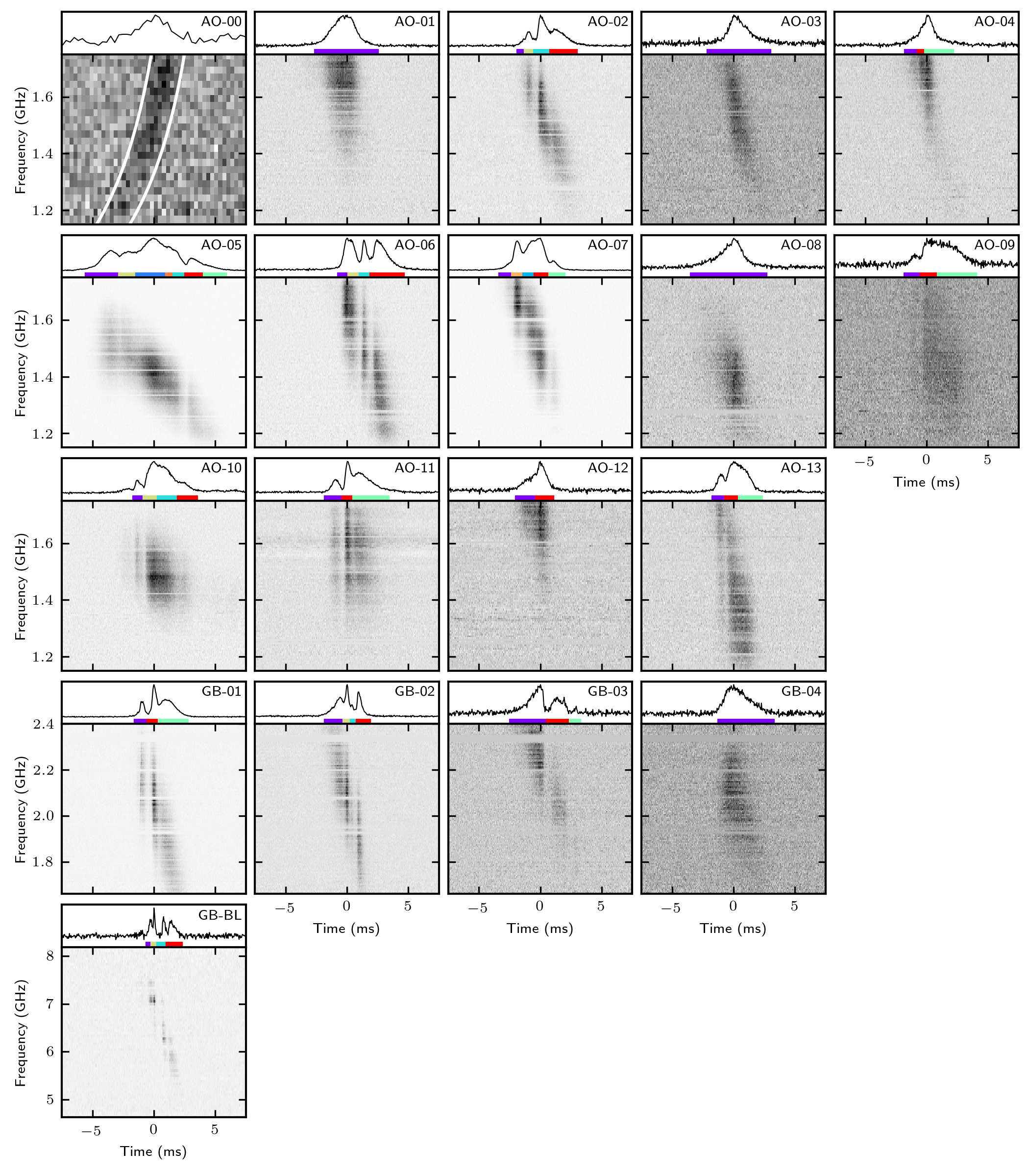}}
\caption{Dynamic spectra of the bursts (see Table~\ref{tab:bursts}), each dedispersed to ${\rm DM} = 560.5$\,pc\,cm$^{-3}$, and using a linear scaling in arbitrary units (the bursts are not flux calibrated).
The plotted dynamic spectra have the following time/frequency resolutions: AO-00: 0.33\,ms/25\,MHz; AO-01-13 and GB-01-04: 0.041\,ms/6.25\,MHz; GB-BL: 0.041\,ms/55.66\,MHz.  The narrow horizontal stripes are the result of flagging RFI-contaminated channels.  At the top of each panel, the band-integrated burst profile is shown, with the colored bars indicating the time spans of the sub-bursts used in the fitting.  Bursts AO-01 to AO-13 are the new bursts detected with Arecibo.  For comparison, AO-00 is burst \#17 from \citet{ssh+16b}; the white lines show the best-fit ${\rm DM} = 559$\,pc\,cm$^{-3}$ for that burst, which deviates significantly from the ${\rm DM} = 560.5$\,pc\,cm$^{-3}$ dispersive correction displayed here.  GB-01 to GB-04 are the four new GBT bursts detected at 2.0\,GHz, and GB-BL is one of the 6.5-GHz GBT Breakthrough Listen bursts presented in \citet{gsp+18}.
\label{fig:weird_bursts}}
\end{figure*}
\begin{figure*}
\centerline{\includegraphics[width=0.99\textwidth]{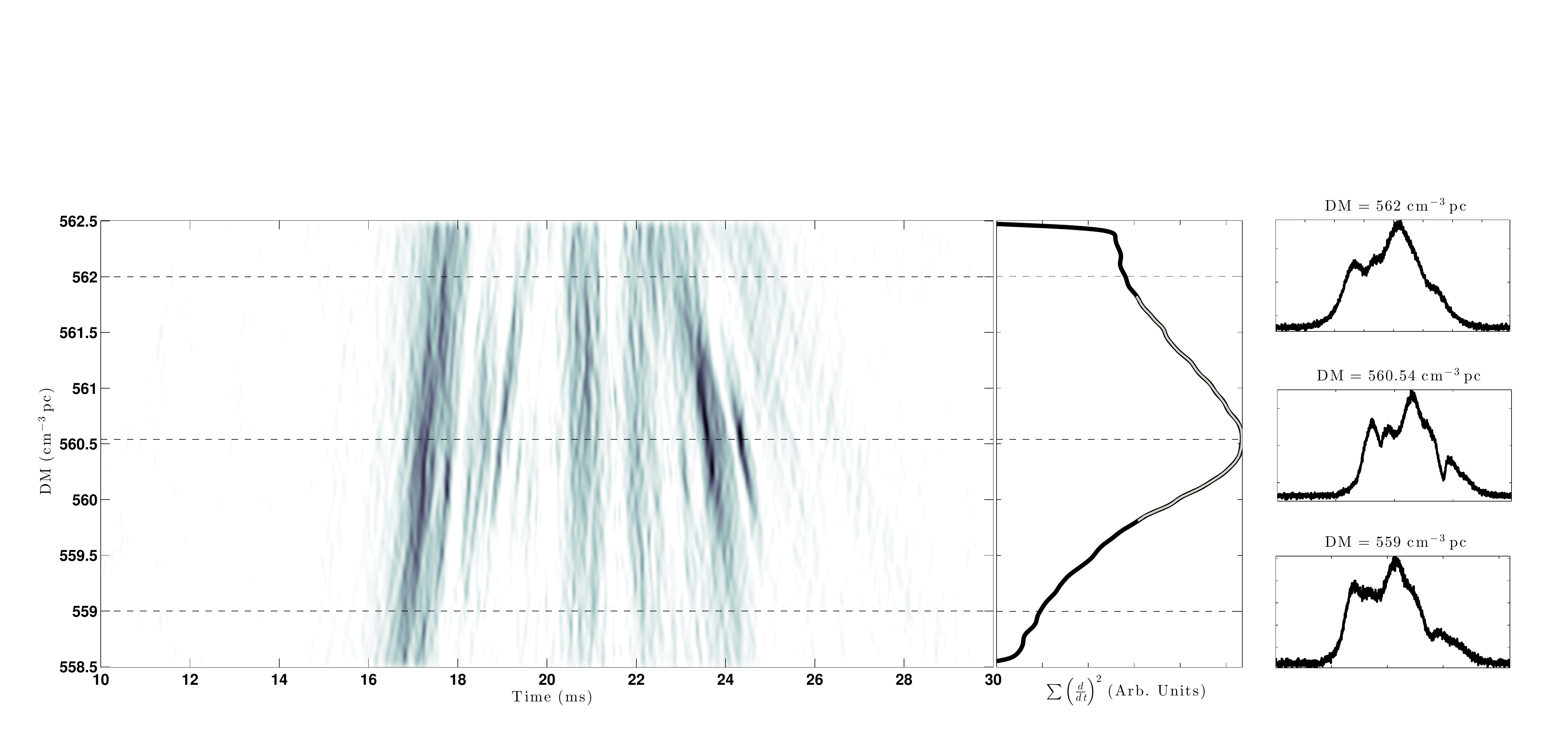}}
\caption{An example of the DM optimization method, using burst AO-05.  The main panel presents the square of the Gaussian-smoothed forward-difference time-derivative of the frequency-averaged burst profile as a function of DM and time.  Darker regions show steeper areas of the profile when varying DM.  The adjacent sub-panel shows the average along the time axis.  Here the gray curve  overlaid on the time-average curve is the high-order polynomial used for the optimal DM interpolation. The right-hand panels show the frequency-averaged burst profiles at DM values above, at, and below the optimum value, which are marked with dash lines in the main panel.
\label{fig:burst_DM}}
\end{figure*}

As can be seen in Figure~\ref{fig:weird_bursts}, the burst sample displays a significantly more complex structure than previously reported bursts from \frb, most of which appeared single peaked \citep{ssh+16a,ssh+16b,sbh+17,msh+18,gsp+18}. In the sample here, bursts show as many as seven components that can be isolated in time and frequency, and which we refer to as sub-bursts. The sub-burst separations are $\sim 1$\,ms, and hence much more closely spaced compared to the shortest published burst separations to date: $\sim 40$\,ms \citep{sbh+17} and 34\,ms \citep{hds+17}.  Though there is typically a gradual rise into the first sub-burst, it often appears that the leading edges of subsequent sub-bursts show a sharper rise in brightness compared to the more gradual decay in the trailing edges.  Shorter-timescale sub-burst structure is sometimes seen on top of wider, more diffuse emission.  Between sub-bursts, there are sometimes sharp drops in brightness.  The overall time-frequency structure is reminiscent of a diffraction pattern, showing isolated peaks and troughs in brightness.  There is no obvious similarity in the time-frequency structures of bursts detected within a single observation, or even for bursts separated by only a few minutes in time.  Of the bursts presented here, the shortest and longest separations between bursts observed within the same observing session are $\sim 138$\,s for bursts AO-01 and AO-02 and $\sim 2360$\,s for bursts AO-11 and AO-12, respectively (Table~\ref{tab:bursts}).  In the following, we quantitatively characterize the burst features.


First, we manually identified individual sub-bursts, whose time spans are indicated by colored bars under the frequency-averaged profiles in Figure~\ref{fig:weird_bursts}.  This is an imperfect time division of the bursts because some sub-bursts are less distinct than others, and because there is sometimes also more diffuse underlying emission.  We used a least-squares fitting routine to measure the characteristic bandwidth and duration of each sub-burst using a 2D Gaussian function.  These Gaussians were aligned along the time and frequency axes, and thus we did not fit for any residual time-frequency drift within sub-bursts.  This is because any such analysis is additionally complicated by frequency evolution of the sub-burst profiles.
Also, we note that this fitting is not significantly influenced by RFI excision, which only affects the spectrum on a much narrower frequency scale compared to the bandwidths of the sub-bursts.

\begin{figure*}
\centerline{\includegraphics[width=0.5\textwidth]{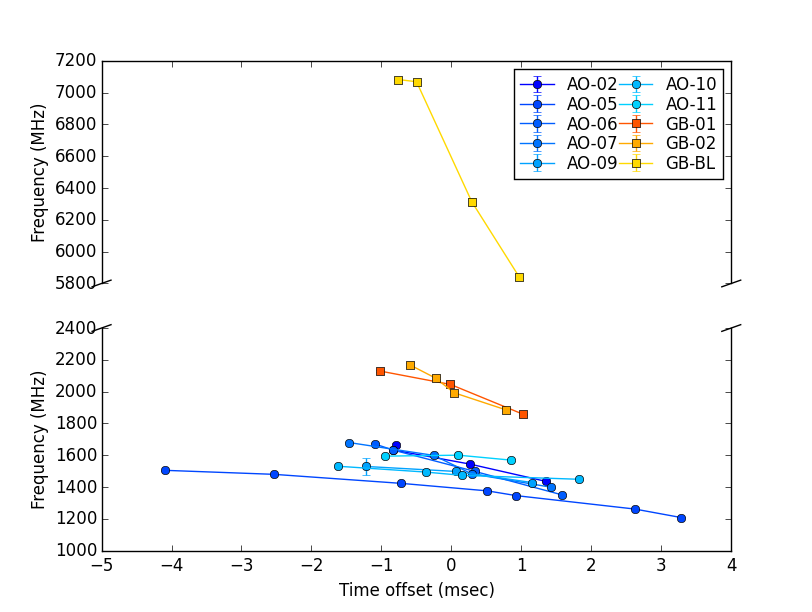} \includegraphics[width=0.5\textwidth]{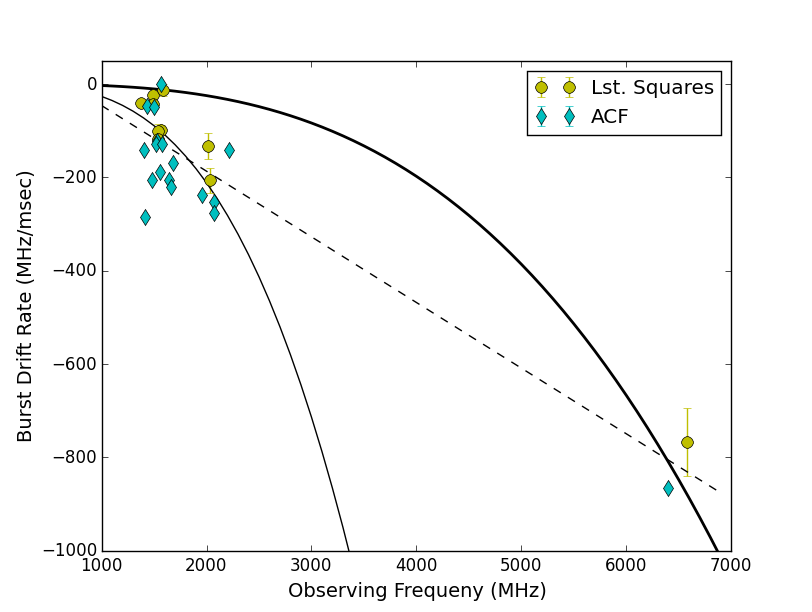}}\centerline{\includegraphics[width=0.5\textwidth]{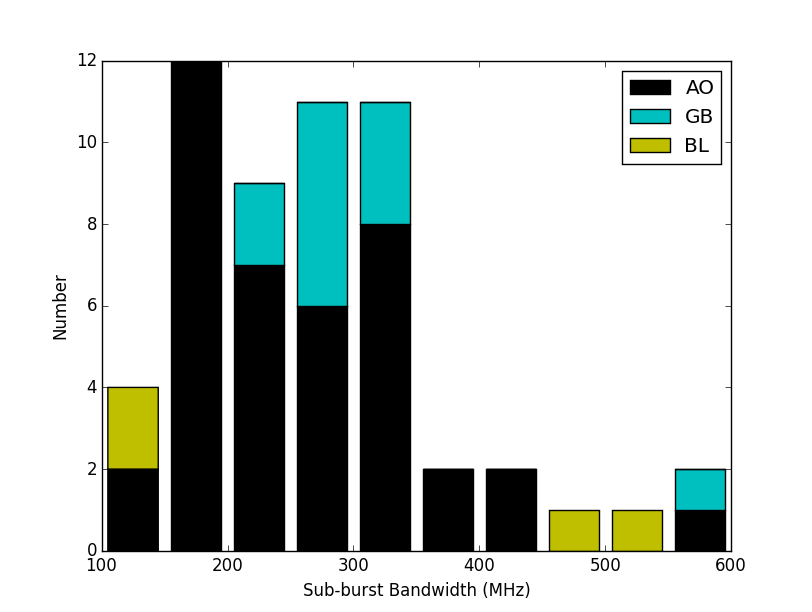} \includegraphics[width=0.5\textwidth]{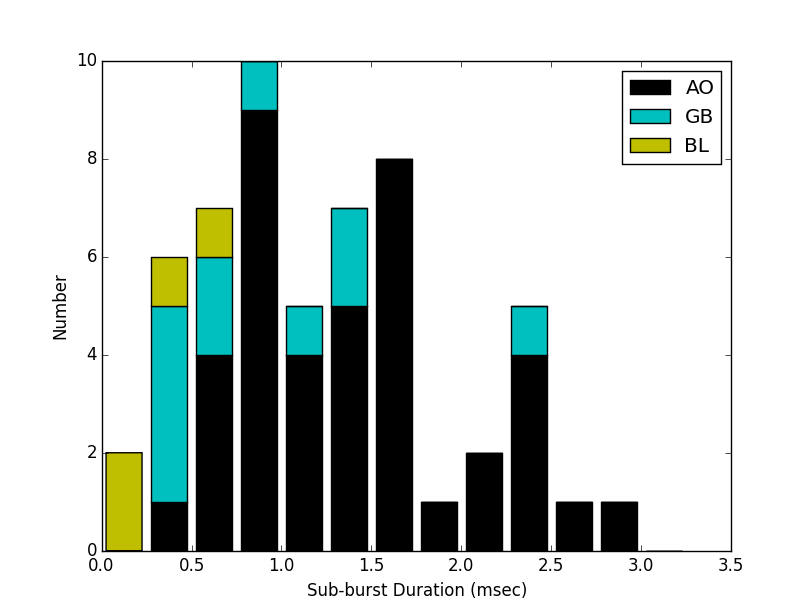}}
\caption{{\it Top-left}: Sub-burst central frequency as a function of arrival time.  The bursts are aligned such that, for each burst, emission at the average frequency of the sub-bursts arrives at zero time offset.  This is to demonstrate that they have similar slopes at the same central observing frequency.  {\it Top-right}: Measured linear burst drift rates versus burst characteristic radio frequency for the least squares (yellow circles) and ACF (cyan diamonds) methods. The solid curves illustrate the drift expected if the DM used to dedisperse the burst was too low. The thicker solid line corresponds to a $\Delta$DM $\sim$40~pc\,cm$^{-3}$ as determined through a least squares fit to all of the data points, while the thiner solid line corresponds to $\Delta$DM $\sim$5~pc\,cm$^{-3}$ as determined through a least squares fit to only the 1.4- and 2.0-GHz bursts. The dashed line illustrates a linear fit to the data.  {\it Bottom-left}: The FWHM bandwidths measured by fitting a 2D Gaussian model to each sub-burst in the sample using the least squares routine. The 1.4-GHz Arecibo bursts are shown in black, the 2.0-GHz GBT bursts in cyan, and the 6.5-GHz GBT bursts in yellow.  {\it Bottom-right}: The 2D Gaussian FWHM temporal durations of each sub-burst as determined by the least squares fitting technique. Color coding same as for {\it Bottom-left}.
\label{fig:burst_fits}}
\end{figure*}

Figure~\ref{fig:burst_fits} shows the distribution of sub-burst bandwidths and durations for the 1.4-GHz, 2.0-GHz and 6.5-GHz bursts. For the 1.4-GHz bursts, we find that the sub-bursts emit with a characteristic bandwidth of $\sim 250$\,MHz, although with a 1-$\sigma$ variation of $\sim 90$\,MHz. For the few 2.0-GHz and 6.5-GHz bursts included in this sample, the characteristic bandwidth is comparable, but somewhat higher on average. Note that the $\sim 100$-MHz features seen in the GB-BL 6.5-GHz sub-bursts are consistent with originating from Galactic diffractive interstellar scintilliation \citep[DISS;][]{gsp+18}.

Overall burst durations at 1.4\,GHz --- defined as the FWHM of the full-burst envelope --- are typically $\sim 3$\,ms and consistent with previous measurements in the literature \citep[e.g.,][]{ssh+16a,ssh+16b}.  However, most bursts show narrower internal structure (sub-bursts) with widths $\lesssim 1$\,ms.  Note that these sub-bursts are resolved in time and are not significantly affected by intra-channel dispersive smearing or interstellar scattering (see \S\ref{sec:scint}). 

Burst durations at 2.0, 4.5, and 6.5\,GHz appear to be systematically smaller than at 1.4\,GHz \citep[see also Fig.~7 of][]{gsp+18}. For example, \citet{msh+18} found total burst durations of $\lesssim 1$\,ms for their sample of bursts detected at 4.5\,GHz.  However, the sample sizes are small and this trend requires confirmation.  Also, these multi-frequency bursts were observed at different epochs, and it is possible that burst width also changes with time, systematically.

\begin{figure*}
\centerline{\includegraphics[width=0.95\textwidth]{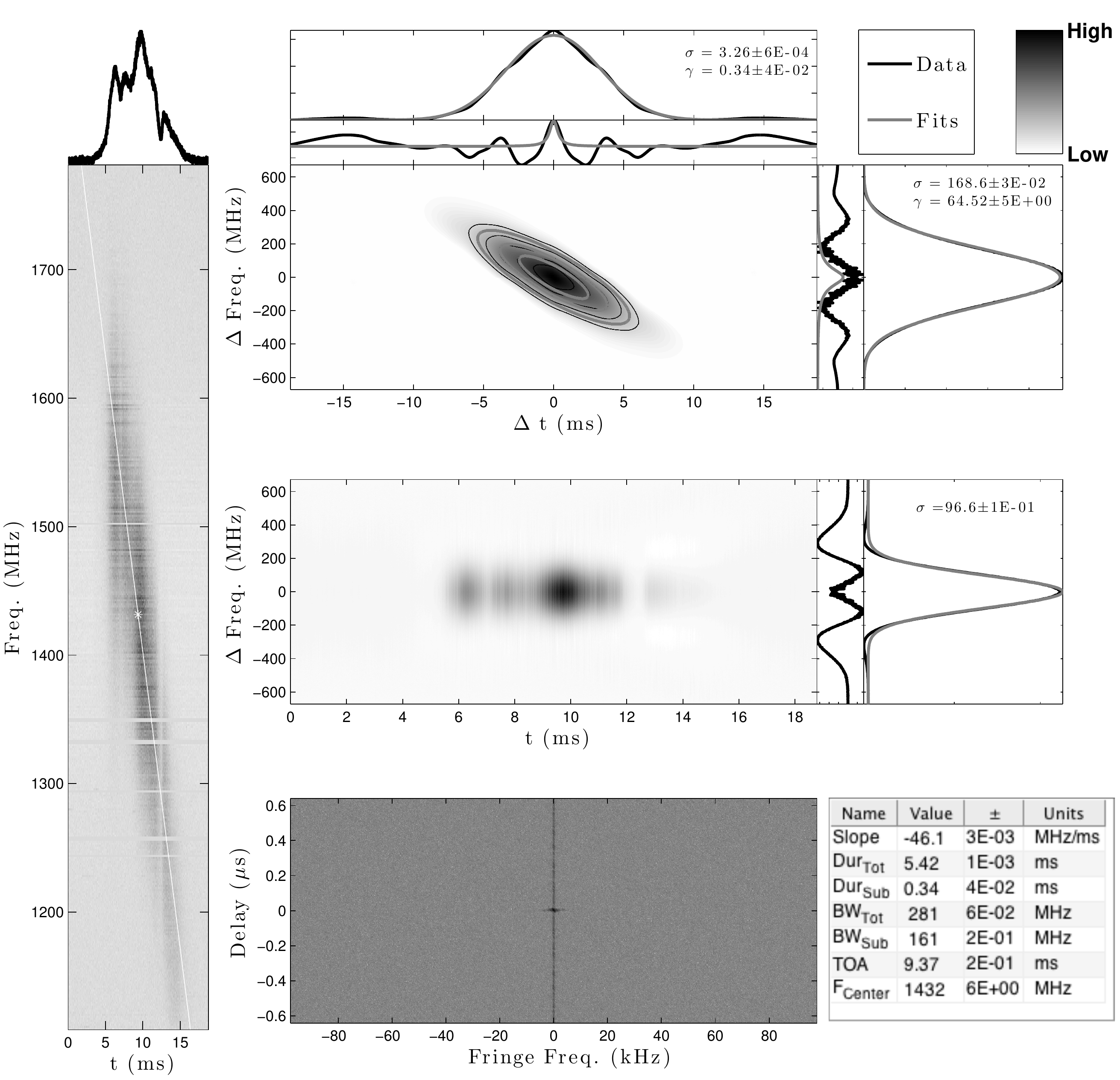}}
\caption{A diagnostic plot from the autocorrelation function (ACF) burst analysis, using burst AO-05 as an example.  {\it Left}: the dynamic spectrum, with the profile averaged over frequency shown above.  Here the white diagonal line and star show the fitted drift and characteristic frequency of the burst.  {\it Top-right}: a two dimensional ACF for the burst, with adjacent sub-panels showing the average along each axis. These average ACF curves are fitted with a Gaussian distribution, and the residuals of those are fitted with a Lorentz distribution.  {\it Center-right}: the non-normalized ACF at each time stamp, with the time-averaged ACF shown in the adjacent sub-panel.  This time-averaged ACF is fitted with a Gaussian, whose residual is displayed. {\it Bottom-right}: the secondary spectrum and a table of fitted values.
\label{fig:burst_fits_corr}}
\end{figure*}

To complement the 2D Gaussian least-squares fitting of individual sub-bursts (which were first manually identified to provide initial parameters to the fit), we also performed an unguided 2D auto-correlation function (ACF) analysis (Figure~\ref{fig:burst_fits_corr}) of the de-dispersed dynamic spectra of the bursts.  The characteristic sub-burst durations 
($W_{\rm sb}$ in Table~\ref{tab:bursts}) are from this analysis.

Particularly striking is the tendency for the characteristic frequency of the sub-bursts (i.e. the central frequency of a band-limited sub-burst) to drift to lower frequencies at later times during the burst.  We characterized this drift using both fitting methods. For the least-squares technique, the centers of the best-fit 2D Gaussians in frequency and time for each burst (Figure~\ref{fig:burst_fits}, Top Left) were fit to a linear model.  Only bursts with three or more components and with frequency centers within the band were included. The resulting slopes are shown in Figure~\ref{fig:burst_fits} (Top Right, yellow circles). Drift rates were also estimated using the ACF method and are listed in Table~\ref{tab:bursts} and shown in Figure~\ref{fig:burst_fits} (Top Right, cyan diamonds). Note that the ACF method has the advantage that it can be applied to all bursts, regardless of their number of components. The inferred ACF drift rates are in good agreement with those derived by fitting the central times and frequencies of individual sub-bursts.

Interestingly, the drift rates of this burst sample are always negative (sub-bursts peak in brightness at lower frequencies at later times), and the magnitude of the drift rate increases with increasing radio frequency.  
In one case, however, AO-05 (Figure~\ref{fig:weird_bursts}), the first two sub-bursts show no drift with respect to each other, and only thereafter does the downward trend begin.

The metric that is used to determine DM is a crucial consideration in interpreting these drifts (see \S\ref{sec:dm_amb}); we would also find a drift to lower frequencies at later times if we were under-dedispersing the bursts: $d\nu/dt \propto -\nu^3/\rm \delta DM$, where $\delta$DM is the residual DM.  We calculated the best-fit $\delta$DM to the estimated drift rates with the GB-BL burst ($\delta$DM $\sim$ 5\,pc\,cm$^{-3}$) and without the GB-BL burst ($\delta$DM $\sim$ 40\,pc\,cm$^{-3}$). These fits are shown in Figure~\ref{fig:burst_fits} as the thick and thin solid lines. Clearly, no single value of $\delta$DM fits the measurements at all three observing frequencies, and we argue that the drift rate is not caused by residual dispersion.  Finally, we fit a line to the rates, and while it is a good fit, the absence of bursts in our sample between $\sim 2$ and 6\,GHz makes any conclusive statement difficult.  Note that the 4.5-GHz bursts presented by \citet{msh+18} do not show any clear examples of sub-burst drift to include in the analysis here.  For that sample, the observing bandwidth of 800\,MHz is comparable to the $\sim 500$\,MHz/ms drift rate that we would predict based on the sample presented here.  In fact, the clear drift visible in the 6.5-GHz GB-BL burst presented here is only visible because of the very large bandwidth of those observations.

\subsection{Scintillation}
\label{sec:scint} 

We argue here that Galactic diffractive interstellar scintillation (DISS) accounts for fine structure in burst spectra but not for
the relatively broadband ($\sim 100-400$\,MHz) frequency structure observed in the 1.4 and 2.0-GHz sub-bursts. 

\begin{figure*}
\centerline{\includegraphics[width=0.48\textwidth]{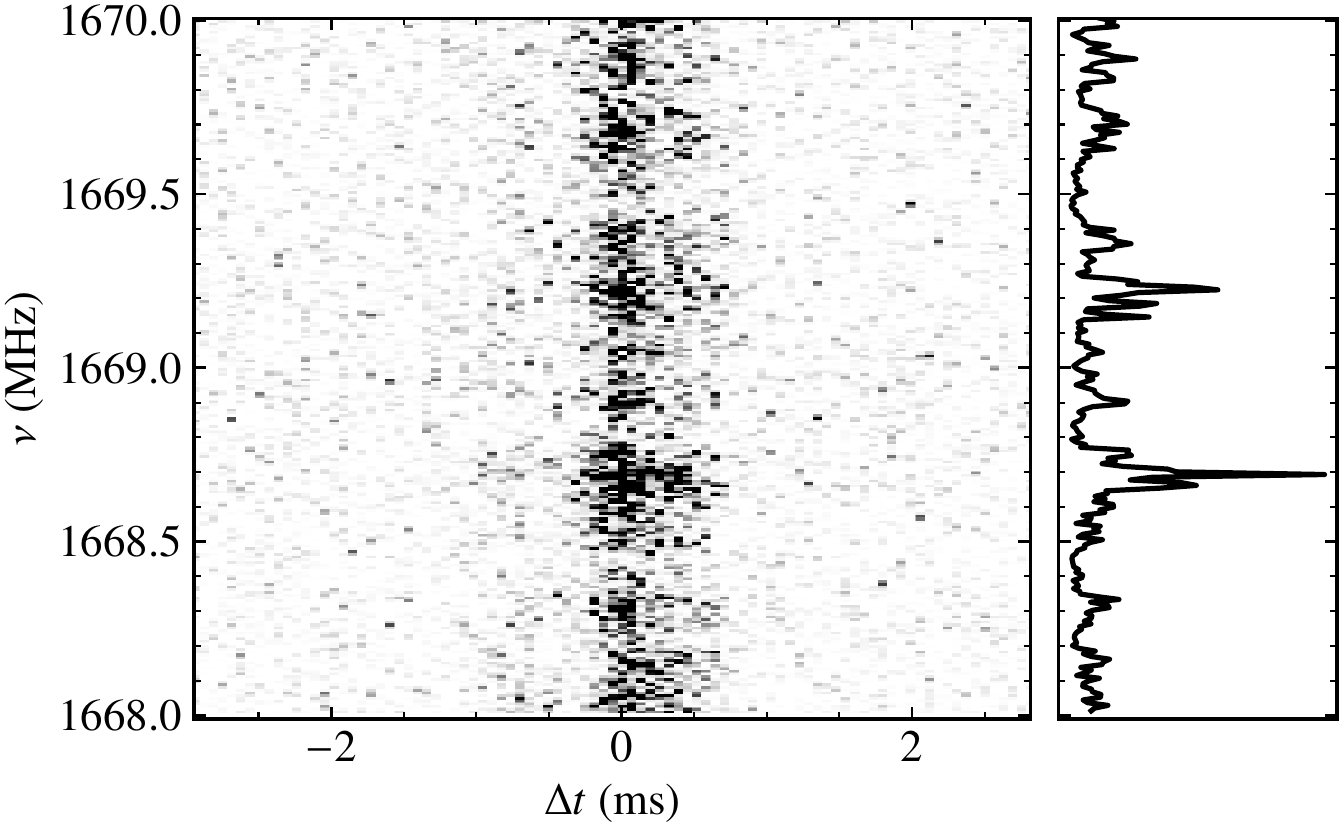}
\hspace{0.5cm}
\includegraphics[width=0.45\textwidth]{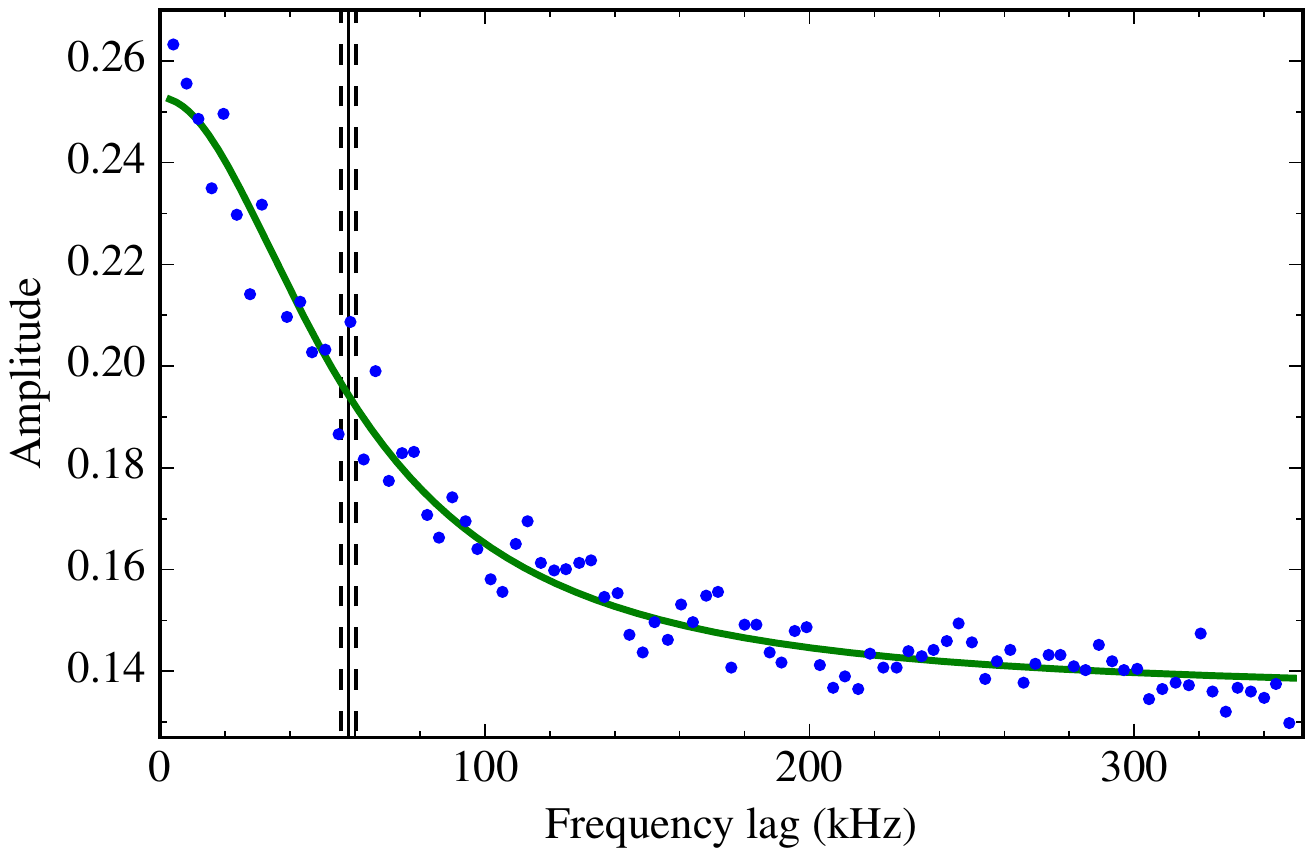}}
\caption{{\it Left}: A zoom-in on 2\,MHz of the dedispersed dynamic spectrum of a burst detected in European VLBI Network (EVN) observations.  The right-hand sub-panel shows the cumulative burst brightness (arbitrary units) as a function of frequency.  {\it Right}: Autocorrelation function of the burst spectrum showing that its narrow-band frequency structure has a characteristic scale (half width at half maximum, HWHM) of $58.1 \pm 2.3$\,kHz.  Here the solid vertical line shows the HWHM of the fitted Lorentzian function (shown by the solid green curve), and the dashed lines show the uncertainty.  
\label{fig:DISS}}
\end{figure*}

To demonstrate this, we re-analyze the brightest European VLBI Network (EVN) burst presented by \citet{mph+17}, using just the auto-correlations from Arecibo.   These voltage data provide only 64\,MHz of spectral coverage, but offer the opportunity for much better frequency resolution compared to the PUPPI/GUPPI data available for the other bursts.  The EVN burst shows that there is fine-scale frequency structure ($< \mathrm{MHz}$) in the total intensity
(Figure~\ref{fig:DISS}).
In principle the structure could be due to DISS exclusively, or a combination between DISS and `self noise' in the pulsar signal.  Burst electric fields are well described as an intrinsic shot-noise process modulated by an envelope function. The resulting spectrum has frequency structure with widths equal to the reciprocal burst width; this structure may then combine with the extrinsically imposed scintillation modulation \citep[][]{cbh+04}.  The self-noise frequency structure is on a much different scale compared to the sub-burst spectral peaks displayed in Figure~\ref{fig:weird_bursts}.  

To measure a characteristic bandwidth for these narrow-band spectral features, we used an ACF analysis \citep{cwb85}.  We computed the ACFs from   power spectral densities generated with a resolution of 3.9\,kHz from the de-dispersed EVN Arecibo voltage data using only the time range that coincides with the burst.  We fitted a Lorentzian function to the ACF using a least-squares approach as implemented in the Levenberg-Marquardt algorithm.  The central lag of the ACF, which is dominated by noise, was excluded from the fit.  Furthermore, because of bandpass effects, only the central 80\% of the frequency range in each of the 4 subbands was used to compute the ACF.  We measure a characteristic bandwidth of $58.1 \pm 2.3$\,kHz at 1.65\,GHz, which corresponds to the half width at half maximum (HWHM) of the fitted Lorentzian function (Figure~\ref{fig:DISS}).  

The characteristic bandwidth
is consistent to better than a factor of two with the NE2001 Galactic electron model prediction for the DISS contribution from the Milky Way in this direction \citep{cl02}: Scaling the  model prediction to 1.65\,GHz using $\nu^4$ and $\nu^{4.4}$, respectively, yields bandwidths of 87 and 107\,kHz. We note that the YMW16 model \citep{ymw17} under predicts the DISS bandwidth by a factor of 30 (1.5\,kHz at 1.65\,GHz); this will be discussed in a separate paper. 

The  pulse broadening time at 1.65\,GHz corresponding to the DISS bandwidth is $(2\pi\times 58.1~{\rm kHz})^{-1} = 2.7~\mu$s, which is much smaller than the time resolution of our data.  The scintillation time scale is unmeasurable because it is expected to be much larger (order of hours) than the burst durations. 

We thus conclude that the narrow ($<$ MHz) frequency structures seen in the bursts are due to DISS imparted when they enter the Galaxy, and consequently that the broad ($\sim 100-400$\,MHz) spectral features and the temporal structure seen in Figure~\ref{fig:weird_bursts} must either be intrinsic or imparted in the local environment of the source (or perhaps elsewhere in \frb's host galaxy).

Similarly, at higher frequencies of $4-8$\,GHz, \citet{gsp+18} and \citet{shb+18} found $5-100$\,MHz frequency structure, which they attributed to Galactic DISS, and which is also consistent with the NE2001 predictions.  This implies that the $\sim 1$\,GHz frequency structure in the 6.5-GHz GB-BL burst presented here is also likely intrinsic or imparted near the source.

\section{Discussion} \label{sec:disc}

\subsection{Comparison of \frb\ with Other FRBs}

\frb\ differs notably from other FRBs in the fact that it repeats in an easily detectable way \citep{ssh+16a}.  The bursts also display an extreme Faraday rotation \citep{msh+18} that has not been seen in any other FRB to date \citep[see Fig.~5 in][which summarizes all available measurements]{cks+18}.  While some FRBs have a reasonably high absolute RM ($|{\rm RM}| \sim 200$\,rad\,m$^{-2}$) that originates close to the source \citep[e.g.][]{mls+15a}, others show a very low absolute RM \citep[$|{\rm RM}| \lesssim 10$\,rad\,m$^{-2}$, e.g.][]{rsb+16}.  However, previous polarimetric FRB detections lacked sufficient frequency resolution to resolve such a large RM as seen in \frb, and hence some FRBs with no apparent linear polarization may have very large RMs as well \citep{pbb+15}.  

Despite the possibility that \frb\ has a fundamentally different origin (or inhabits a markedly different environment) compared to the apparently non-repeating FRBs, it is nonetheless useful to compare its burst structure to what has been seen in other FRBs.  The repeating nature and localization of \frb\ have allowed higher time- and frequency-resolution data to be acquired over a relatively large range of frequencies.  As such, the detailed time-frequency features it displays may foreshadow what other FRBs will show in similar observations.

While \frb\ bursts can clearly be multi-peaked, the majority of non-repeating FRB bursts detected to date appear simple in form.  However, in some cases this may simply be because they are broadened by uncorrected intra-channel dispersion smearing \citep{rav18} or by scattering \citep{tsb+13} --- either of which can mask sub-millisecond temporal structure.  The  multi-component FRB~121002 and FRB~130729 show time-frequency structures similar to those of \frb\, albeit at lower S/N \citep{cpk+16}, though the unknown position of these bursts with respect to the telescope sensitivity pattern makes it difficult to interpret their spectra.  

More recently, \citet{ffb+18} present the UTMOST discovery of FRB~170827 at a central observing frequency of 835\,MHz.
Three temporal components, one only $\sim 30$\,$\mu$s wide, were detected in FRB~170827's burst profile thanks to real-time triggering of voltage data, which allowed coherent dedispersion.  With the coarser time sampling, and incoherent dedispersion used to discover this source, this same burst looks similar to the single-component FRBs detected with Parkes \citep{pbj+16}.  This suggests that other high-S/N FRBs analyzed with coherent dedispersion will also show complex structure.  The narrow bandwidth (31\,MHz) available in the detection of FRB~170827 limits the ability to see whether its sub-bursts drift in frequency like \frb.  The data also do not allow for an RM measurement.  Regardless, the burst time structure and timescales are similar to those of \frb.  One can thus speculate that, despite FRB~170827's apparent non-repeatability \citep{ffb+18}, this suggests a similar physical origin to \frb\ and provides an observational bridge between the Repeater and non-repeating FRBs. Bright FRBs in the ASKAP sample also appear to show some spectral structure \citep{smb+18}. While they have only a single temporal component, perhaps due to limited time resolution of the observations, several bursts show large spectral modulation. \citet{msb+18} attribute these to propagation effects since they observe an anti-correlation between modulation and DM. These new observations further highlight that complex spectra appear to be relevant for apparently non-repeating FRBs as well.

\subsection{Comparison with Radio Emission from Neutron Stars}

Based on light-travel-time arguments, the short durations of FRB pulses require compact emission regions.  For example, the 30-$\mu$s-wide component detected in one \frb\ pulse requires an emitting region $\lesssim 10$\,km, assuming no additional geometric or relativistic effects \citep{msh+18}.  Thus it is natural to compare FRB emission to neutron star radio emission, even though \frb\ has thus far shown no clear periodicity in its burst arrival times \citep{ssh+16a,zgf+18}.  Like \frb, pulsars and magnetars show a wide range of pulse complexity in the time domain.  This results from the rotation of fluctuating beamed radiation across the line-of-sight.  \frb\ differs markedly from pulsars and magnetars in several ways, however; in particular, its bursts are enormously more energetic.  Both pulsar pulses and FRBs have peak flux densities $\sim 1$\,Jy but the $\sim 10^6$ times greater distance of \frb\ implies a $\sim 10^{12}$ times greater luminosity (for equal solid angles), corresponding to burst energies

\be
\label{eq:burst_energy}
E_{\rm burst} = 4\pi D^2 (\delta\Omega/4\pi) A_{\nu} \Delta\nu 
\approx 10^{38}\, {\rm erg}\,(\delta\Omega/4\pi) D_{\rm Gpc}^2  (A_{\nu} / 0.1\ {\rm Jy\ ms}) \Delta\nu_{\rm GHz},
\ee
where $\delta\Omega$ is the emission solid angle in steradians, $D_{\rm Gpc}$ the luminosity distance in Gpc, $A_{\nu}$ the fluence in Jy~ms, and $\Delta\nu_{\rm GHz}$ the emission bandwidth in GHz (in the source frame).  Pulsar-type magnetospheres may have difficulty in providing this energy \citep[e.g.][]{cw16,lyut17}.  Alternatively, bursts from \frb\ may be powered by the strong $\sim 10^{14} - 10^{15}$\,G magnetic fields in magnetars \citep{pp13,bel17}.  

Another marked difference between \frb\ and typical pulsars and radio-emitting magnetars is in the spectral domain, where the latter objects have smooth, wide-band spectra \citep[even in their single pulses, e.g.,][]{kkg+03,jsk+18} whose only narrow-band modulation is from DISS, augmented in some cases by constructive and destructive interference from multiple imaging by interstellar refraction.  While the radio-emitting magnetars have shown variable spectra, these remain well fit by a broad-band power law \citep[e.g.,][]{ljk+08}.
In contrast, the confinement of \frb\ bursts to frequency bands of width $\sim 250$\,MHz (at $\sim 1.4$\,GHz) is different compared to variable magnetar spectra, and also cannot be explained by Galactic DISS.  To our knowledge, no similar effect is seen in pulsars except for the high-frequency interpulse of the Crab pulsar, or in cases of plasma lensing (which we will discuss in the following sub-section).

Indeed, the giant pulse emission in the Crab pulsar's high-frequency interpulse \citep[HFIP;][]{hej16}, seen at radio frequencies above $\sim 4$\,GHz, provides an intriguing observational analogy.   
Notably, the properties of the HFIPs differ significantly from those of the main giant pulses \citep[MP;][]{jpk+10,hej16}.  Since the Crab is a young ($\sim$1000 year old) neutron star embedded in a luminous nebula, it is also an interesting Galactic example of the young PWN/SNR scenario for \frb.  It is possible that the \frb\ system is simply a much younger version of the Crab, though understanding the scaling to the energies required by \frb\ remains challenging.  As Eq.~\ref{eq:burst_energy} shows, a highly focused beam, or intrinsically narrow-band emission can reduce the required energy.

The Crab's HFIP spectra exhibit periodic banded structure \citep{he07} with separations $\Delta\nu$ that scale with frequency ($\Delta\nu/\nu = $ constant). Drift rates in \frb\ may show a similar scaling (Figure~\ref{fig:burst_fits}) but there are too few bursts in our sample to be conclusive.  Furthermore, we note that while the Crab HFIPs are microseconds in duration, the burst envelopes of \frb\ are typically milliseconds --- though with underlying $\sim 30$\,$\mu$s structure clearly visible in some cases \citep{msh+18}.  Searches for even finer-timescale structure in \frb\ should continue, using high observing frequencies to avoid smearing from scattering.
 
The polarization angle of the $\sim 100$\% linearly polarized radiation from \frb\ at $4-8$\,GHz appears constant across bursts and is stable between bursts \citep{msh+18,gsp+18}. Here again there is phenomenological similarity with the Crab's HFIPs, which are $\sim 80-100$\% linearly polarized with a constant polarization position angle across the duration of each pulse and also between HFIPs that span $\sim 3$\% of the pulsar's rotational phase \citep[see Fig.~14 of][]{hej16}.  Like \frb, the Crab HFIPs typically also show no circular polarization. 

\subsection{Intrinsic Processes and Propagation Effects}
 
The spectral properties of \frb\ may be intrinsic to the radiation process, post-emission propagation processes, or some combination of the two.  

Spectral structure is seen in bursts from the Sun \citep[e.g.,][]{kmi+15}, flare stars \citep[e.g.,][]{ob06,ob08}, and Solar System planets \citep[e.g.][]{Zar92,rzh+14}, including auroral kilometric radiation from the Earth and Saturn and the decametric radiation from Jupiter \citep[e.g.,][]{tre06}.  Frequency drifts, qualitatively similar to those seen from \frb, occur due to upward motions of emission regions to locations with smaller plasma frequencies or cyclotron frequencies, which are tied to the observed electromagnetic frequency.  Fine structure in the emission is related to structure in the particle density \citep[e.g.,][]{tre06}.  Extrapolation of similar processes to FRBs suggests that \frb's emission could originate from cyclotron or synchrotron maser emission \citep{lyub14,bel17,wax17}, in which case relatively narrow-band emission in the GHz range could be expected. Antenna mechanisms involving curvature radiation from charge bunches have also been considered
\citep{cw16,lk17} but it is not clear if the energetics can be satisfied or how time-frequency structure is produced. 

Alternatively, burst propagation through media outside the emission region can also produce spectral features by refraction and diffraction from large- and small-scale structure in ionized plasma, respectively.  Enhanced electron densities in confined regions can act as diverging (overdensities) or converging (underdensities) lenses - i.e. `plasma lenses'.  The resulting effects produce highly chromatic amplifications and multiple images \citep{cfl98,cwh+17} with bandwidths strongly dependent on the detailed properties of the lenses.  Multiple images of bursts will have different amplitudes, peak frequencies, arrival times, and DMs.   If  burst images overlap in time and frequency, they can produce interference structure on small time and frequency scales, including oscillations that follow the square of an  Airy function \citep{wm06,cwh+17}.  This is qualitatively similar to what we observe from \frb, and though we can model individual bursts well with a single DM, small differences ($\lesssim 0.1$\,pc\,cm$^{-3}$) in DM between sub-bursts may still be present, allowing for the possibility of different bursts being slightly differently lensed.

\citet{msh+18} argue that \frb\ is embedded in a compact, ionized region with a magnetic field of at least a few milli-Gauss and a substantial electron density ($n_e \gtrsim 10$\,cm$^{-3}$). 
The large RM suggests that the ionized gas is dominated by a non-relativistic Hydrogen-Helium plasma because a relativistic gas or gas comprising an electron-positron plasma would yield a small or null RM. 

The large variation in RM between bursts separated by 7 months --- without a similarly large accompanying DM variation --- indicates that the region is dynamic, possibly much smaller than 1\,pc in thickness, and contains even smaller $\sim$AU-size structures that could cause plasma lensing.  Depending on the ratio of thermal to magnetic pressure in the plasma, $\beta$, and the geometry of the field (disordered or misaligned from the line-of-sight), the requirements for plasma lensing give a consistent picture for the measured RM if the region's depth is of order $\sim$AU, the electron density $\sim 10^4$\,cm$^{-3}$ and the field $\gtrsim 1$\,mG.  Note that the magnetic field strength could even be a thousand times larger, $\sim 1$\,G, if the DM related to the Faraday region is small ($\lesssim 1$\,pc\,cm$^{-3}$).  

The detection of transient pulse echoes from the Crab pulsar presents an observational precedent for plasma lensing \citep{glj11}.  While these echoes are fainter than the normal Crab pulsar emission, the possibility that \frb\ is also embedded in a dense nebula suggests an interesting analogy.  Though such large RMs as seen from \frb\ have not been observed in the Crab pulses, the Crab echo events are associated with apparent DM variations\footnote{These variations are much larger and rapid compared to the $10^{-2} - 10^{-4}$\,pc\,cm$^{-3}$ variations seen over year-long timescales along normal pulsar lines of sight through the Galactic interstellar medium \citep{hlk+04}.} of $\sim 0.1$\,pc\,cm$^{-3}$ \citep{bwv00}, which are similar but less extreme compared to the order $\sim 1$\,pc\,cm$^{-3}$ variations seen in \frb.

More recently, \citet{myc+18} discovered that plasma lensing can boost the observed brightness of the `black widow' Galactic millisecond pulsar PSR~B1957+20 in a strongly time and frequency-dependent way\footnote{Similar effects have also been seen in PSR~B1744$-$24A \citep{brn11}.}.  PSR~B1957+20 is a binary millisecond pulsar, which is eclipsed by intra-binary material blown off of the companion star by the pulsar wind.  The plasma lensing events occur near eclipse ingress and egress, and last for a few to tens of milliseconds.  Their dynamic spectra \citep[see Fig.~2 of][]{myc+18} are qualitatively similar to those of \frb\ presented here.  While this is a stunning demonstration of how plasma lensing can boost the observed brightness of pulsed radio emission by close to two orders-of-magnitude, we note that \frb\ likely inhabits a much different environment compared with PSR~B1957+20 \citep{msh+18}.

Furthermore, while plasma lensing can explain the downwards frequency drift of the \frb\ sub-pulses, this would require a single dominant lens for the drift to be in the same direction for some amount of time.  If plasma lensing is the cause for the sub-burst frequency drift, one would expect the drift rate to change rate and sign with time, as  the viewing geometry changes and different lenses dominate.  In the case of PSR~B1957+20, where many lenses are involved, brightness enhancements are seen to drift both upwards and downwards over the course of tens of milliseconds \citep{myc+18}.

\subsection{Constraints on the Magneto-ionic Medium Near \frb}

\newcommand{\DMRMhundred}{\DM_{\RM, 100}}

Here we assume that all of the source-frame rotation measure, ${\rm RM}_{\rm src} = 1.46 \times 10^5~\RMunits$, is from a thin region near the source, with thickness $l$. However,  the associated $\DMRM$ from the same region may be substantially less than that inferred for the host galaxy, $\DMhost \approx 100~\DMunits$. 
Constraints on the thickness, magnetization, and temperature of the Faraday region can be derived from the observations \citep[see also][]{msh+18}, and extended further under the assumption that plasma lensing is responsible for the observed time-frequency structures in \frb\ bursts.

We relate the parallel magnetic field estimated from ${\rm RM}_{\rm src}$ to the magnetic pressure and obtain the electron density
\be
n_e &=& 
	4.6\times 10^4 ~{\rm cm^{-3}} \,
    \DMRMhundred^{-2}
	\left(\frac{\eta_B^2 T_4}{\beta}\right)^{-1},
\label{eq:ne1}
\ee
and region thickness
\be
l &=& 
	449~{\rm AU} \times  \DMRMhundred^{3}
	\left(\frac{\eta_B^2 T_4}{\beta}\right),
\label{eq:l1}
\ee
where $\DMRMhundred$ is the DM associated with the Faraday medium in units of $100\,\DMunits$, $T_4$ is the temperature in units of $10^4$\,K, $\beta$ is a scaling factor between the magnetic field and thermal energy densities ($\beta = 1$ in the case of equipartition), and
$\eta_B \le 1$ accounts either for the misalignment of an ordered magnetic field with respect to the line of sight or for a turbulent field with local values much larger than the net parallel component that contributes to the RM.   We refer to the combination $\eta_B^2 T_4/ \beta$ as the composite gas factor.  These in turn imply a free-free optical depth
\be
\tau_{\rm ff} 
&\approx &
 1.5 \, T_4^{-1.3} \, \nu^{-2.1} \, \DMRMhundred^{-1}
	\left(\frac{\eta_B^2 T_4}{\beta}\right)^{-1}.
\label{eq:tauff}
\ee
If the DM in the Faraday region is small, e.g. $\DMRM = 1~\DMunits$, the implied optical depth is large even at 1~GHz unless the temperature alone or the composite gas factor $\eta_B^2 T_4 / \beta$ are also large. 

If plasma lensing is relevant, then there is 
an additional constraint. Equation~7 of Cordes et al. (2017) gives the focal distance for a plasma lens, which is the distance required for ray crossing at a frequency $\nul$ where lensing has been seen (or inferred).    Requiring the focal distance   to be less than the $\sim1$~Gpc distance of FRB~121102 yields a constraint involving the lens dispersion measure, $\DM_l$, 
\be
\frac{(\aau\nul)^2}{\DM_l \dsl} \le 1.5\, {\dso},
\label{eq:lenslim1}
\ee
where $\aau$ is the transverse radius of the lens, $\dsl$ is the source-lens distance in pc,   the source-observer distance (in Gpc) is  $\dso \sim 1$, and $\nul$ is in GHz.  We define the depth of the lens $l$  (along the line of sight) to be a  multiple $A$ of the lens diameter $2\aau$, i.e. $\lau = 2A\aau$. 

An upper bound on the depth is then
\be
l \le 24.5\,{\rm AU}~ (\dsl\dso)^{1/2} 
\DMRMhundred^{1/2}
\left(\frac{A}{\nul}  \right). 
\label{eq:l_upper_bound}
\ee
Combining Eq.~\ref{eq:l1} and \ref{eq:l_upper_bound} gives an upper bound on the gas factor
\be
\left(\frac{\eta_B^2 T_4}{\beta}\right) 
	\le 
	0.055  \left(\frac{A}{\nul}\right)   
    \frac{(\dsl\dso)^{1/2}}{\DMRMhundred^{5/2}}.
\label{eq:gas_factor_upper_bound}
\ee
This in turn gives a lower bound on the electron density
\be
n_{\rm e} 
	\ge 
	8.42\times 10^5 ~{\rm cm^{-3}}\, 
    \left(\frac{\DMRMhundred} {\dso\dsl}  \right)^{1/2}
	 \left(\frac{\nul}{A}\right).
\label{eq:ne_lower_bound}
\ee
The lower bound on electron density yields a lower bound on the  free-free optical depth,
\be
\tau_{\rm ff}  &\ge& 
 28.4 \, T_4^{-1.3} \, \nu^{-2.1}
 	\, \DMRMhundred^{3/2}
 	\left(\dso\dsl \right)^{-1/2}
	 \left(\frac{\nul}{A}\right).
 \label{eq:tauff_lower_bound}
\ee

Equations~\ref{eq:l_upper_bound} to \ref{eq:tauff_lower_bound} comprise the constraints that  can be made from the RM measurements combined with constraints on DM from H$\alpha$ measurements \citep{tbc+17,bta+17,kms+17} and from the interpretation of burst time-frequency structure as being the result of plasma lensing.  

Clearly the  free-free optical depth is far too large for nominal values of $\DMRM$ and $T_4$.  If the dispersion depth is instead much smaller than 100~$\DMunits$ or if the temperature is much greater than $10^4$~K, the optical depth can be reduced below unity at 1 GHz.   A reduced $\DMRM$ also makes the Faraday region thinner and the magnetic field larger.   It also increases the upper bound on the gas factor substantially owing to the strong dependence $\propto \DMRM^{5/2}$. 

If multiple imaging from lensing causes the time-frequency structure in bursts from FRB~121102 at frequencies as high as about 8~GHz (as the GB-BL burst presented here suggests as a possibility), then $\nul = 8$~GHz.  This increases all derived quantities proportionately except for the gas parameter, which is reduced by the same factor.  

Overall there appears to be sufficient latitude to account for the measured Faraday rotation as well as for the requirements for plasma lensing to be met. For a small $\DMRM = 1\, \DMunits$,  
 the Faraday region is very thin ($l \lesssim 1$-10~AU),  
 highly magnetized ($B \gtrsim 1$~G),
  and has a high electron density ($n_e \gtrsim 10^5$~cm$^{-3}$).  Intriguingly, these values are comparable to those inferred for the Crab echo events, where \citet{glj11} argued that these are created by plasma lensing from filaments with diameters of $\sim 2$\,AU and electron density of the order of $10^4$\,cm$^{-3}$.

For $\DMRM = 1\, \DMunits$ the optical depth need not be large, $\tau_{\rm ff} \gtrsim 0.03$ at 1~GHz,  but will be optically thick at frequencies no lower than about 100~MHz.    Of course since we have only a lower bound on $\tau_{\rm ff}$, the region could be optically thick at substantially higher frequencies.

The apparent increase of $\sim 1-3$\,pc\,cm$^{-3}$ in \frb's DM over 4 years could indicate a genuine increase in electron column density along the line-of-sight, e.g. from an expanding supernova shock-wave sweeping up ambient material \citep{pg18}.  However, we again caution that this is not necessarily a secular trend, and it could also reflect frequency-dependent arrival time delays due to variable plasma lensing like seen in the Crab \citep{bwv00}.

\subsection{Consequences for Observing \frb\ and Other FRBs}

The discovery of detailed time-frequency structures in the bursts from \frb\ has various general consequences for FRB searches as well as follow-up observations aimed at detecting repeat bursts.

First, more sophisticated detection algorithms, like 2D matched filtering, should be considered in order to blindly identify bursts with complicated frequency and time structure (e.g., Zackay et al., in prep.).  The traditional method of collapsing the entire frequency band and searching with a bank of 1D boxcar matched filters is sub-optimal if (sub-)bursts occupy only a fraction of the observing band.  Also, machine learning algorithms can help detect FRBs with unusual time-frequency structure that would otherwise be missed or discarded as RFI \citep{zgf+18}.

Furthermore, care is needed in measuring and interpreting the DMs of bursts.  Evolving burst morphology with radio frequency can bias the DM to be high (or low) if a simple metric of peak S/N optimization is used.  While this effect will likely only subtly change the apparent DM, application of an erroneous DM can smooth-out temporal structure in a band-integrated burst profile.  Also, when searching for repeat bursts from other FRBs, coherent dedispersion should be used in order to maintain sensitivity to potential sub-millisecond bursts that could be missed if the dispersive smearing is too large.  

If plasma lensing significantly affects \frb's burst brightness, then this can explain why its detectability changes with time.  This is relevant for constraining the repeatability of other FRB sources, and the spectra of other FRBs should also be examined for evidence of plasma lensing.  

Lastly, identifying any underlying periodicity in the bursts from \frb\ caused, for example, by rotation of a compact object, is complicated by the burst structures and by any variable propagation delays that exceed the putative rotational period.  This issue is greatest when the underlying period is not much longer than the $\sim 1-10$\,ms envelopes of the bursts themselves.  This also suggests that burst detections at higher frequencies, where propagation delays are expected to be significantly lower, may be more fruitful for identifying an underlying periodicity in the burst arrival times.  However, such efforts have, as yet, been unsuccessful \citep{zgf+18}.

\section{Conclusions and Future Work} \label{sec:conc}

We have shown that radio bursts detected from \frb\ often exhibit complex time-frequency structure that is unlike what is commonly seen in radio pulsars or radio-emitting magnetars.  We apply a DM determination metric that maximizes structure in the frequency-averaged pulse profile, and which reveals that bursts are composed of temporally distinct sub-bursts with widths $\lesssim 1$\,ms and characteristic emission bandwidths of typically $\sim 250$\,MHz at $\sim 1.4$\,GHz.  Furthermore, these sub-bursts drift to lower frequencies with time at a rate of $\sim 200$\,MHz/ms at 1.4\,GHz, and the rate of drift is possibly larger at higher radio frequencies.  We find that the bursts in this sample have a ${\rm DM} = 560.57 \pm 0.07$\,pc\,cm$^{-3}$ at MJD~57644, and this suggests an increase of $\Delta{\rm DM} \sim 1-3$\,pc\,cm$^{-3}$ in 4 years.  Whether this is a smooth, secular increase or whether there are stochastic variations at the $\sim 1$\,pc\,cm$^{-3}$ level is, as yet, unclear.

We have discussed how the time-frequency structures in the bursts could be intrinsic to the emission mechanism, or due to local propagation effects.  While the \frb\ bursts show many commonalities with the Crab pulsar high-frequency interpulses, the time-frequency structures are also consistent with plasma lensing, like that seen from the Crab nebula and in the intra-binary material of PSR~B1957+20.  In either case, the time-frequency structure provides new information about the nature of the underlying bursting source and its environment.  Overall, these new findings are consistent with previously proposed scenarios in which \frb\ is a particularly young neutron star in a dense nebula.

A larger, high-S/N, and broad frequency burst sample is needed to further address the nature of \frb.  In the absence of prompt multi-wavelength counterpart, the radio bursts themselves remain a key diagnostic.  Future work can better quantify DM variations, whether the apparent drift rate of the sub-bursts changes with time, and whether there is a correlation between the variable RM and the time-frequency structure in the bursts.  If the RM is dominated by a single plasma lens, correlated variations could be expected.  Furthermore, a larger sample can address if sub-burst brightness is inversely proportional to its characteristic bandwidth and whether individual sub-bursts have demonstrably different DMs --- both of which would be expected in a plasma lensing scenario.  Continued monitoring, over the broadest possible range of radio frequencies, and preferentially with simultaneous ultra-broadband observations, is thus strongly motivated.

The low frequencies and huge fractional bandwidth ($400-800$\,MHz) offered by CHIME \citep{chime18} is well suited to exploring the role of local propagation effects like plasma lensing --- especially if bursts can be studied in fine detail using coherent dedispersion on buffered voltage data.  While \frb\ has yet to be detected below 1\,GHz \citep{ssh+16b}, both UTMOST and CHIME have shown that FRBs are detectable at these frequencies \citep{ffb+18,boy18}.
Finding commonalities in the burst properties between repeating and apparently non-repeating FRBs may establish that they have common physical origins but different activity levels or more/less fortuitous viewing geometries with respect to their beamed emission.  

\newpage

\acknowledgments
We thank Ue-Li Pen for interesting discussions and comments on an early draft of this paper.  We thank the staff of the Arecibo Observatory, Green Bank Observatory, and European VLBI Network for their continued support and dedication to enabling observations like those presented here.  The Arecibo Observatory is operated by SRI International under a cooperative agreement with the National Science Foundation (AST-1100968), and in alliance with Ana G.~M\'{e}ndez-Universidad Metropolitana, and the Universities Space Research Association.  The Green Bank Observatory is a facility of the National Science Foundation (NSF) operated under cooperative agreement by Associated Universities, Inc.  The European VLBI Network is a joint facility of independent European, African, Asian, and North American radio astronomy institutes.  Scientific results from data presented in this publication are derived from the following EVN project codes: RP024 and RP026.  J.W.T.H. and D.M. acknowledge funding from an NWO Vidi fellowship and from the European Research Council (ERC) under the European Union's Seventh Framework Programme (FP/2007-2013) / ERC Starting Grant agreement nr. 337062 (``DRAGNET'').  L.G.S. acknowledges financial support from the ERC Starting Grant BEACON (nr. 279702), as well as the Max Planck Society.  S.C., J.M.C., F.C., M.A.M., and S.M.R. acknowledge support from the NANOGrav Physics Frontiers Center, funded by the National Science Foundation (NSF) award number 1430284.  S.C. and J.M.C. acknowledge support from the NSF award AAG 1815242.  M.A.M. is also supported by NSF award number OIA-1458952.  A.M.A. is an NWO Veni fellow.
J.S.D. was supported by the NASA Fermi program.  V.G. acknowledges NSF grant 1407804 and the Marilyn and Watson Alberts SETI Chair funds.  V.M.K. is supported by a Lorne Trottier Chair in Astrophysics \& Cosmology, a Canada Research Chair, by an NSERC Discovery Grant and Herzberg Award, by FRQNT/CRAQ and is a Senior CIFAR Fellow.  C.J.L. is supported by NSF award 1611606.  B.M. acknowledges support from the Spanish Ministerio de Econom\'{i}a y Competitividad (MINECO) under grants AYA2016-76012-C3-1-P and MDM-2014-0369 of ICCUB (Unidad de Excelencia ``Mar\'{i}a de Maeztu'').  S.M.R. is a CIFAR Senior Fellow.  P.S. is supported by a DRAO Covington Fellowship from the National Research Council Canada.    L.C. and E.P. acknowledge funding from ERC grant nr. 617199.  B.W.S. acknowledges funding from the ERC under the European Union's Horizon 2020 research and innovation programme (nr. 694745).  

%

\vspace{5mm}
\facilities{Arecibo, GBT, EVN}


\software{Astropy, DSPSR, PSRCHIVE, PRESTO, psrfits\_utils}




\bibliographystyle{aasjournal}
\bibliography{weird_bursts}

\begin{thebibliography}{}
\expandafter\ifx\csname natexlab\endcsname\relax\def\natexlab#1{#1}\fi
\providecommand{\url}[1]{\href{#1}{#1}}

\bibitem[{{Backer} {et~al.}(2000){Backer}, {Wong}, \& {Valanju}}]{bwv00}
{Backer}, D.~C., {Wong}, T., \& {Valanju}, J. 2000, \apj, 543, 740

\bibitem[{{Bassa} {et~al.}(2016){Bassa}, {Beswick}, {Tingay}, {Keane},
  {Bhandari}, {Johnston}, {Totani}, {Tominaga}, {Yasuda}, {Stappers}, {Barr},
  {Kramer}, \& {Possenti}}]{bbt+16}
{Bassa}, C.~G., {Beswick}, R., {Tingay}, S.~J., {et~al.} 2016, \mnras, 463, L36

\bibitem[{{Bassa} {et~al.}(2017){Bassa}, {Tendulkar}, {Adams}, {Maddox},
  {Bogdanov}, {Bower}, {Burke-Spolaor}, {Butler}, {Chatterjee}, {Cordes},
  {Hessels}, {Kaspi}, {Law}, {Marcote}, {Paragi}, {Ransom}, {Scholz},
  {Spitler}, \& {van Langevelde}}]{bta+17}
{Bassa}, C.~G., {Tendulkar}, S.~P., {Adams}, E.~A.~K., {et~al.} 2017, \apjl,
  843, L8

\bibitem[{{Beloborodov}(2017)}]{bel17}
{Beloborodov}, A.~M. 2017, \apjl, 843, L26

\bibitem[{{Bilous} {et~al.}(2011){Bilous}, {Ransom}, \& {Nice}}]{brn11}
{Bilous}, A.~V., {Ransom}, S.~M., \& {Nice}, D.~J. 2011, in American Institute
  of Physics Conference Series, Vol. 1357, American Institute of Physics
  Conference Series, ed. M.~{Burgay}, N.~{D'Amico}, P.~{Esposito},
  A.~{Pellizzoni}, \& A.~{Possenti}, 140--141

\bibitem[{{Boyle} \& {Chime/Frb Collaboration}(2018)}]{boy18}
{Boyle}, P.~C., \& {Chime/Frb Collaboration}. 2018, The Astronomer's Telegram,
  11901

\bibitem[{{Caleb} {et~al.}(2018){Caleb}, {Keane}, {van Straten}, {Kramer},
  {Macquart}, {Bailes}, {Barr}, {Bhat}, {Bhandari}, {Burgay}, {Farah},
  {Jameson}, {Jankowski}, {Johnston}, {Petroff}, {Possenti}, {Stappers},
  {Tiburzi}, \& {Venkatraman Krishnan}}]{cks+18}
{Caleb}, M., {Keane}, E.~F., {van Straten}, W., {et~al.} 2018, \mnras, 478,
  2046

\bibitem[{{Champion} {et~al.}(2016){Champion}, {Petroff}, {Kramer}, {Keith},
  {Bailes}, {Barr}, {Bates}, {Bhat}, {Burgay}, {Burke-Spolaor}, {Flynn},
  {Jameson}, {Johnston}, {Ng}, {Levin}, {Possenti}, {Stappers}, {van Straten},
  {Thornton}, {Tiburzi}, \& {Lyne}}]{cpk+16}
{Champion}, D.~J., {Petroff}, E., {Kramer}, M., {et~al.} 2016, \mnras, 460, L30

\bibitem[{{Chatterjee} {et~al.}(2017){Chatterjee}, {Law}, {Wharton},
  {Burke-Spolaor}, {Hessels}, {Bower}, {Cordes}, {Tendulkar}, {Bassa},
  {Demorest}, {Butler}, {Seymour}, {Scholz}, {Abruzzo}, {Bogdanov}, {Kaspi},
  {Keimpema}, {Lazio}, {Marcote}, {McLaughlin}, {Paragi}, {Ransom}, {Rupen},
  {Spitler}, \& {van Langevelde}}]{clw+17}
{Chatterjee}, S., {Law}, C.~J., {Wharton}, R.~S., {et~al.} 2017, \nat, 541, 58

\bibitem[{{CHIME/FRB Collaboration} {et~al.}(2018){CHIME/FRB Collaboration},
  {Amiri}, {Bandura}, {Berger}, {Bhardwaj}, {Boyce}, {Boyle}, {Brar},
  {Burhanpurkar}, {Chawla}, {Chowdhury}, {Cliche}, {Cranmer}, {Cubranic},
  {Deng}, {Denman}, {Dobbs}, {Fandino}, {Fonseca}, {Gaensler}, {Giri},
  {Gilbert}, {Good}, {Guliani}, {Halpern}, {Hinshaw}, {H{\"o}fer}, {Josephy},
  {Kaspi}, {Landecker}, {Lang}, {Liao}, {Masui}, {Mena-Parra}, {Naidu},
  {Newburgh}, {Ng}, {Patel}, {Pen}, {Pinsonneault-Marotte}, {Pleunis}, {Rafiei
  Ravandi}, {Ransom}, {Renard}, {Scholz}, {Sigurdson}, {Siegel}, {Smith},
  {Stairs}, {Tendulkar}, {Vanderlinde}, \& {Wiebe}}]{chime18}
{CHIME/FRB Collaboration}, {Amiri}, M., {Bandura}, K., {et~al.} 2018, \apj,
  863, 48

\bibitem[{{Clegg} {et~al.}(1998){Clegg}, {Fey}, \& {Lazio}}]{cfl98}
{Clegg}, A.~W., {Fey}, A.~L., \& {Lazio}, T.~J.~W. 1998, \apj, 496, 253

\bibitem[{{Connor} {et~al.}(2016){Connor}, {Sievers}, \& {Pen}}]{csp16}
{Connor}, L., {Sievers}, J., \& {Pen}, U.-L. 2016, \mnras, 458, L19

\bibitem[{{Cordes} {et~al.}(2004){Cordes}, {Bhat}, {Hankins}, {McLaughlin}, \&
  {Kern}}]{cbh+04}
{Cordes}, J.~M., {Bhat}, N.~D.~R., {Hankins}, T.~H., {McLaughlin}, M.~A., \&
  {Kern}, J. 2004, \apj, 612, 375

\bibitem[{{Cordes} \& {Lazio}(2002)}]{cl02}
{Cordes}, J.~M., \& {Lazio}, T.~J.~W. 2002, ArXiv Astrophysics e-prints,
  astro-ph/0207156

\bibitem[{{Cordes} \& {Wasserman}(2016)}]{cw16}
{Cordes}, J.~M., \& {Wasserman}, I. 2016, \mnras, 457, 232

\bibitem[{{Cordes} {et~al.}(2017){Cordes}, {Wasserman}, {Hessels}, {Lazio},
  {Chatterjee}, \& {Wharton}}]{cwh+17}
{Cordes}, J.~M., {Wasserman}, I., {Hessels}, J.~W.~T., {et~al.} 2017, \apj,
  842, 35

\bibitem[{{Cordes} {et~al.}(1985){Cordes}, {Weisberg}, \& {Boriakoff}}]{cwb85}
{Cordes}, J.~M., {Weisberg}, J.~M., \& {Boriakoff}, V. 1985, \apj, 288, 221

\bibitem[{{Cordes} {et~al.}(2006){Cordes}, {Freire}, {Lorimer}, {Camilo},
  {Champion}, {Nice}, {Ramachandran}, {Hessels}, {Vlemmings}, {van Leeuwen},
  {Ransom}, {Bhat}, {Arzoumanian}, {McLaughlin}, {Kaspi}, {Kasian}, {Deneva},
  {Reid}, {Chatterjee}, {Han}, {Backer}, {Stairs}, {Deshpande}, \&
  {Faucher-Gigu{\`e}re}}]{cfl+06}
{Cordes}, J.~M., {Freire}, P.~C.~C., {Lorimer}, D.~R., {et~al.} 2006, \apj,
  637, 446

\bibitem[{{DeLaunay} {et~al.}(2016){DeLaunay}, {Fox}, {Murase},
  {M{\'e}sz{\'a}ros}, {Keivani}, {Messick}, {Mostaf{\'a}}, {Oikonomou}, {Te{\v
  s}i{\'c}}, \& {Turley}}]{dfm+16}
{DeLaunay}, J.~J., {Fox}, D.~B., {Murase}, K., {et~al.} 2016, \apjl, 832, L1

\bibitem[{{DuPlain} {et~al.}(2008){DuPlain}, {Ransom}, {Demorest}, {Brandt},
  {Ford}, \& {Shelton}}]{drd+08}
{DuPlain}, R., {Ransom}, S., {Demorest}, P., {et~al.} 2008, in \procspie, Vol.
  7019, Advanced Software and Control for Astronomy II, 70191D

\bibitem[{{Eatough} {et~al.}(2013){Eatough}, {Falcke}, {Karuppusamy}, {Lee},
  {Champion}, {Keane}, {Desvignes}, {Schnitzeler}, {Spitler}, {Kramer},
  {Klein}, {Bassa}, {Bower}, {Brunthaler}, {Cognard}, {Deller}, {Demorest},
  {Freire}, {Kraus}, {Lyne}, {Noutsos}, {Stappers}, \& {Wex}}]{efk+13}
{Eatough}, R.~P., {Falcke}, H., {Karuppusamy}, R., {et~al.} 2013, \nat, 501,
  391

\bibitem[{{Farah} {et~al.}(2018){Farah}, {Flynn}, {Bailes}, {Jameson},
  {Bannister}, {Barr}, {Bateman}, {Bhandari}, {Caleb}, {Campbell-Wilson},
  {Chang}, {Deller}, {Green}, {Hunstead}, {Jankowski}, {Keane}, {Macquart},
  {M{\"o}ller}, {Onken}, {Os{\l}owski}, {Parthasarathy}, {Ravi}, {Shannon},
  {Tucker}, {Venkatraman Krishnan}, \& {Wolf}}]{ffb+18}
{Farah}, W., {Flynn}, C., {Bailes}, M., {et~al.} 2018, ArXiv e-prints,
  arXiv:1803.05697

\bibitem[{{Gajjar} {et~al.}(2017){Gajjar}, {Siemion}, {MacMahon}, {Croft},
  {Hellbourg}, {Isaacson}, {Enriquez}, {Price}, {Lebofsky}, {DeBoer},
  {Werthimer}, {Hickish}, {Brinkman}, {Chatterjee}, \& {Ransom}}]{gsm+17}
{Gajjar}, V., {Siemion}, A.~P.~V., {MacMahon}, D.~H.~E., {et~al.} 2017, The
  Astronomer's Telegram, 10675

\bibitem[{{Gajjar} {et~al.}(2018){Gajjar}, {Siemion}, {Price}, {Law},
  {Michilli}, {Hessels}, {Chatterjee}, {Archibald}, {Bower}, {Brinkman},
  {Burke-Spolaor}, {Cordes}, {Croft}, {Enriquez}, {Foster}, {Gizani},
  {Hellbourg}, {Isaacson}, {Kaspi}, {Lazio}, {Lebofsky}, {Lynch}, {MacMahon},
  {McLaughlin}, {Ransom}, {Scholz}, {Seymour}, {Spitler}, {Tendulkar},
  {Werthimer}, \& {Zhang}}]{gsp+18}
{Gajjar}, V., {Siemion}, A.~P.~V., {Price}, D.~C., {et~al.} 2018, \apj, 863, 2

\bibitem[{{Gould} \& {Lyne}(1998)}]{gl98}
{Gould}, D.~M., \& {Lyne}, A.~G. 1998, \mnras, 301, 235

\bibitem[{{Graham Smith} {et~al.}(2011){Graham Smith}, {Lyne}, \&
  {Jordan}}]{glj11}
{Graham Smith}, F., {Lyne}, A.~G., \& {Jordan}, C. 2011, \mnras, 410, 499

\bibitem[{{Hankins} \& {Eilek}(2007)}]{he07}
{Hankins}, T.~H., \& {Eilek}, J.~A. 2007, \apj, 670, 693

\bibitem[{{Hankins} {et~al.}(2016){Hankins}, {Eilek}, \& {Jones}}]{hej16}
{Hankins}, T.~H., {Eilek}, J.~A., \& {Jones}, G. 2016, \apj, 833, 47

\bibitem[{{Hankins} \& {Rickett}(1975)}]{hr75}
{Hankins}, T.~H., \& {Rickett}, B.~J. 1975, Methods in Computational Physics,
  14, 55

\bibitem[{{Hardy} {et~al.}(2017){Hardy}, {Dhillon}, {Spitler}, {Littlefair},
  {Ashley}, {De Cia}, {Green}, {Jaroenjittichai}, {Keane}, {Kerry}, {Kramer},
  {Malesani}, {Marsh}, {Parsons}, {Possenti}, {Rattanasoon}, \&
  {Sahman}}]{hds+17}
{Hardy}, L.~K., {Dhillon}, V.~S., {Spitler}, L.~G., {et~al.} 2017, \mnras, 472,
  2800

\bibitem[{{Hobbs} {et~al.}(2004){Hobbs}, {Lyne}, {Kramer}, {Martin}, \&
  {Jordan}}]{hlk+04}
{Hobbs}, G., {Lyne}, A.~G., {Kramer}, M., {Martin}, C.~E., \& {Jordan}, C.
  2004, \mnras, 353, 1311

\bibitem[{{Jankowski} {et~al.}(2018){Jankowski}, {van Straten}, {Keane},
  {Bailes}, {Barr}, {Johnston}, \& {Kerr}}]{jsk+18}
{Jankowski}, F., {van Straten}, W., {Keane}, E.~F., {et~al.} 2018, \mnras, 473,
  4436

\bibitem[{{Jessner} {et~al.}(2010){Jessner}, {Popov}, {Kondratiev}, {Kovalev},
  {Graham}, {Zensus}, {Soglasnov}, {Bilous}, \& {Moshkina}}]{jpk+10}
{Jessner}, A., {Popov}, M.~V., {Kondratiev}, V.~I., {et~al.} 2010, \aap, 524,
  A60

\bibitem[{{Johnston} {et~al.}(2017){Johnston}, {Keane}, {Bhandari}, {Macquart},
  {Tingay}, {Barr}, {Bassa}, {Beswick}, {Burgay}, {Chandra}, {Honma}, {Kramer},
  {Petroff}, {Possenti}, {Stappers}, \& {Sugai}}]{jkb+17}
{Johnston}, S., {Keane}, E.~F., {Bhandari}, S., {et~al.} 2017, \mnras, 465,
  2143

\bibitem[{{Kaneda} {et~al.}(2015){Kaneda}, {Misawa}, {Iwai}, {Tsuchiya}, \&
  {Obara}}]{kmi+15}
{Kaneda}, K., {Misawa}, H., {Iwai}, K., {Tsuchiya}, F., \& {Obara}, T. 2015,
  \apjl, 808, L45

\bibitem[{{Kashiyama} \& {Murase}(2017)}]{km17}
{Kashiyama}, K., \& {Murase}, K. 2017, \apjl, 839, L3

\bibitem[{{Keane} {et~al.}(2016){Keane}, {Johnston}, {Bhandari}, {Barr},
  {Bhat}, {Burgay}, {Caleb}, {Flynn}, {Jameson}, {Kramer}, {Petroff},
  {Possenti}, {van Straten}, {Bailes}, {Burke-Spolaor}, {Eatough}, {Stappers},
  {Totani}, {Honma}, {Furusawa}, {Hattori}, {Morokuma}, {Niino}, {Sugai},
  {Terai}, {Tominaga}, {Yamasaki}, {Yasuda}, {Allen}, {Cooke}, {Jencson},
  {Kasliwal}, {Kaplan}, {Tingay}, {Williams}, {Wayth}, {Chandra}, {Perrodin},
  {Berezina}, {Mickaliger}, \& {Bassa}}]{kjb+16}
{Keane}, E.~F., {Johnston}, S., {Bhandari}, S., {et~al.} 2016, \nat, 530, 453

\bibitem[{{Kokubo} {et~al.}(2017){Kokubo}, {Mitsuda}, {Sugai}, {Ozaki},
  {Minowa}, {Hattori}, {Hayano}, {Matsubayashi}, {Shimono}, {Sako}, \&
  {Doi}}]{kms+17}
{Kokubo}, M., {Mitsuda}, K., {Sugai}, H., {et~al.} 2017, \apj, 844, 95

\bibitem[{{Kramer} {et~al.}(2003){Kramer}, {Karastergiou}, {Gupta}, {Johnston},
  {Bhat}, \& {Lyne}}]{kkg+03}
{Kramer}, M., {Karastergiou}, A., {Gupta}, Y., {et~al.} 2003, \aap, 407, 655

\bibitem[{{Law} {et~al.}(2017){Law}, {Abruzzo}, {Bassa}, {Bower},
  {Burke-Spolaor}, {Butler}, {Cantwell}, {Carey}, {Chatterjee}, {Cordes},
  {Demorest}, {Dowell}, {Fender}, {Gourdji}, {Grainge}, {Hessels}, {Hickish},
  {Kaspi}, {Lazio}, {McLaughlin}, {Michilli}, {Mooley}, {Perrott}, {Ransom},
  {Razavi-Ghods}, {Rupen}, {Scaife}, {Scott}, {Scholz}, {Seymour}, {Spitler},
  {Stovall}, {Tendulkar}, {Titterington}, {Wharton}, \& {Williams}}]{lab+17}
{Law}, C.~J., {Abruzzo}, M.~W., {Bassa}, C.~G., {et~al.} 2017, ArXiv e-prints,
  arXiv:1705.07553

\bibitem[{{Lazaridis} {et~al.}(2008){Lazaridis}, {Jessner}, {Kramer},
  {Stappers}, {Lyne}, {Jordan}, {Serylak}, \& {Zensus}}]{ljk+08}
{Lazaridis}, K., {Jessner}, A., {Kramer}, M., {et~al.} 2008, \mnras, 390, 839

\bibitem[{{Lazarus} {et~al.}(2015){Lazarus}, {Brazier}, {Hessels},
  {Karako-Argaman}, {Kaspi}, {Lynch}, {Madsen}, {Patel}, {Ransom}, {Scholz},
  {Swiggum}, {Zhu}, {Allen}, {Bogdanov}, {Camilo}, {Cardoso}, {Chatterjee},
  {Cordes}, {Crawford}, {Deneva}, {Ferdman}, {Freire}, {Jenet}, {Knispel},
  {Lee}, {van Leeuwen}, {Lorimer}, {Lyne}, {McLaughlin}, {Siemens}, {Spitler},
  {Stairs}, {Stovall}, \& {Venkataraman}}]{lbh+15}
{Lazarus}, P., {Brazier}, A., {Hessels}, J.~W.~T., {et~al.} 2015, \apj, 812, 81

\bibitem[{{Lorimer} {et~al.}(2007){Lorimer}, {Bailes}, {McLaughlin},
  {Narkevic}, \& {Crawford}}]{lbm+07}
{Lorimer}, D.~R., {Bailes}, M., {McLaughlin}, M.~A., {Narkevic}, D.~J., \&
  {Crawford}, F. 2007, Science, 318, 777

\bibitem[{{Lorimer} \& {Kramer}(2004)}]{lk04}
{Lorimer}, D.~R., \& {Kramer}, M. 2004, {Handbook of Pulsar Astronomy}

\bibitem[{{Lu} \& {Kumar}(2017)}]{lk17}
{Lu}, W., \& {Kumar}, P. 2017, ArXiv e-prints, arXiv:1710.10270

\bibitem[{{Lyubarsky}(2014)}]{lyub14}
{Lyubarsky}, Y. 2014, \mnras, 442, L9

\bibitem[{{Lyutikov}(2017)}]{lyut17}
{Lyutikov}, M. 2017, \apjl, 838, L13

\bibitem[{{Lyutikov} {et~al.}(2016){Lyutikov}, {Burzawa}, \& {Popov}}]{lbp16}
{Lyutikov}, M., {Burzawa}, L., \& {Popov}, S.~B. 2016, \mnras, 462, 941

\bibitem[{{Macquart} {et~al.}(2018){Macquart}, {Shannon}, {Bannister}, {James},
  {Ekers}, \& {Bunton}}]{msb+18}
{Macquart}, J.-P., {Shannon}, R.~M., {Bannister}, K.~W., {et~al.} 2018, ArXiv
  e-prints, arXiv:1810.04353

\bibitem[{{Main} {et~al.}(2018){Main}, {Yang}, {Chan}, {Li}, {Lin}, {Mahajan},
  {Pen}, {Vanderlinde}, \& {van Kerkwijk}}]{myc+18}
{Main}, R., {Yang}, I.-S., {Chan}, V., {et~al.} 2018, \nat, 557, 522

\bibitem[{{Marcote} {et~al.}(2017){Marcote}, {Paragi}, {Hessels}, {Keimpema},
  {van Langevelde}, {Huang}, {Bassa}, {Bogdanov}, {Bower}, {Burke-Spolaor},
  {Butler}, {Campbell}, {Chatterjee}, {Cordes}, {Demorest}, {Garrett}, {Ghosh},
  {Kaspi}, {Law}, {Lazio}, {McLaughlin}, {Ransom}, {Salter}, {Scholz},
  {Seymour}, {Siemion}, {Spitler}, {Tendulkar}, \& {Wharton}}]{mph+17}
{Marcote}, B., {Paragi}, Z., {Hessels}, J.~W.~T., {et~al.} 2017, \apjl, 834, L8

\bibitem[{{Margalit} {et~al.}(2018){Margalit}, {Metzger}, {Berger}, {Nicholl},
  {Eftekhari}, \& {Margutti}}]{mmb+18}
{Margalit}, B., {Metzger}, B.~D., {Berger}, E., {et~al.} 2018, ArXiv e-prints,
  arXiv:1806.05690

\bibitem[{{Masui} {et~al.}(2015){Masui}, {Lin}, {Sievers}, {Anderson}, {Chang},
  {Chen}, {Ganguly}, {Jarvis}, {Kuo}, {Li}, {Liao}, {McLaughlin}, {Pen},
  {Peterson}, {Roman}, {Timbie}, {Voytek}, \& {Yadav}}]{mls+15a}
{Masui}, K., {Lin}, H., {Sievers}, J., {et~al.} 2015, \nat, 528, ??

\bibitem[{{Metzger} {et~al.}(2017){Metzger}, {Berger}, \& {Margalit}}]{mbm17}
{Metzger}, B.~D., {Berger}, E., \& {Margalit}, B. 2017, \apj, 841, 14

\bibitem[{{Michilli} {et~al.}(2018){Michilli}, {Seymour}, {Hessels}, {Spitler},
  {Gajjar}, {Archibald}, {Bower}, {Chatterjee}, {Cordes}, {Gourdji}, {Heald},
  {Kaspi}, {Law}, {Sobey}, {Adams}, {Bassa}, {Bogdanov}, {Brinkman},
  {Demorest}, {Fernandez}, {Hellbourg}, {Lazio}, {Lynch}, {Maddox}, {Marcote},
  {McLaughlin}, {Paragi}, {Ransom}, {Scholz}, {Siemion}, {Tendulkar}, {van
  Rooy}, {Wharton}, \& {Whitlow}}]{msh+18}
{Michilli}, D., {Seymour}, A., {Hessels}, J.~W.~T., {et~al.} 2018, \nat, 553,
  182

\bibitem[{{Murase} {et~al.}(2016){Murase}, {Kashiyama}, \&
  {M{\'e}sz{\'a}ros}}]{mkm16}
{Murase}, K., {Kashiyama}, K., \& {M{\'e}sz{\'a}ros}, P. 2016, \mnras, 461,
  1498

\bibitem[{{Osten} \& {Bastian}(2006)}]{ob06}
{Osten}, R.~A., \& {Bastian}, T.~S. 2006, \apj, 637, 1016

\bibitem[{{Osten} \& {Bastian}(2008)}]{ob08}
---. 2008, \apj, 674, 1078

\bibitem[{{Petroff} {et~al.}(2015{\natexlab{a}}){Petroff}, {Johnston}, {Keane},
  {van Straten}, {Bailes}, {Barr}, {Barsdell}, {Burke-Spolaor}, {Caleb},
  {Champion}, {Flynn}, {Jameson}, {Kramer}, {Ng}, {Possenti}, \&
  {Stappers}}]{pjk+15}
{Petroff}, E., {Johnston}, S., {Keane}, E.~F., {et~al.} 2015{\natexlab{a}},
  \mnras, 454, 457

\bibitem[{{Petroff} {et~al.}(2015{\natexlab{b}}){Petroff}, {Bailes}, {Barr},
  {Barsdell}, {Bhat}, {Bian}, {Burke-Spolaor}, {Caleb}, {Champion}, {Chandra},
  {Da Costa}, {Delvaux}, {Flynn}, {Gehrels}, {Greiner}, {Jameson}, {Johnston},
  {Kasliwal}, {Keane}, {Keller}, {Kocz}, {Kramer}, {Leloudas}, {Malesani},
  {Mulchaey}, {Ng}, {Ofek}, {Perley}, {Possenti}, {Schmidt}, {Shen},
  {Stappers}, {Tisserand}, {van Straten}, \& {Wolf}}]{pbb+15}
{Petroff}, E., {Bailes}, M., {Barr}, E.~D., {et~al.} 2015{\natexlab{b}},
  \mnras, 447, 246

\bibitem[{{Petroff} {et~al.}(2016){Petroff}, {Barr}, {Jameson}, {Keane},
  {Bailes}, {Kramer}, {Morello}, {Tabbara}, \& {van Straten}}]{pbj+16}
{Petroff}, E., {Barr}, E.~D., {Jameson}, A., {et~al.} 2016, \pasa, 33, e045

\bibitem[{{Piro}(2016)}]{pir16}
{Piro}, A.~L. 2016, \apjl, 824, L32

\bibitem[{{Piro} \& {Gaensler}(2018)}]{pg18}
{Piro}, A.~L., \& {Gaensler}, B.~M. 2018, \apj, 861, 150

\bibitem[{{Popov} \& {Postnov}(2013)}]{pp13}
{Popov}, S.~B., \& {Postnov}, K.~A. 2013, ArXiv e-prints, arXiv:1307.4924

\bibitem[{{Ransom}(2001)}]{ran01}
{Ransom}, S.~M. 2001, PhD thesis, Harvard University

\bibitem[{{Ravi}(2018)}]{rav18}
{Ravi}, V. 2018, \mnras, arXiv:1710.08026

\bibitem[{{Ravi} {et~al.}(2015){Ravi}, {Shannon}, \& {Jameson}}]{rsj15}
{Ravi}, V., {Shannon}, R.~M., \& {Jameson}, A. 2015, \apjl, 799, L5

\bibitem[{{Ravi} {et~al.}(2016){Ravi}, {Shannon}, {Bailes}, {Bannister},
  {Bhandari}, {Bhat}, {Burke-Spolaor}, {Caleb}, {Flynn}, {Jameson}, {Johnston},
  {Keane}, {Kerr}, {Tiburzi}, {Tuntsov}, \& {Vedantham}}]{rsb+16}
{Ravi}, V., {Shannon}, R.~M., {Bailes}, M., {et~al.} 2016, Science, 354, 1249

\bibitem[{{Ryabov} {et~al.}(2014){Ryabov}, {Zarka}, {Hess}, {Konovalenko},
  {Litvinenko}, {Zakharenko}, {Shevchenko}, \& {Cecconi}}]{rzh+14}
{Ryabov}, V.~B., {Zarka}, P., {Hess}, S., {et~al.} 2014, \aap, 568, A53

\bibitem[{{Scholz} {et~al.}(2016){Scholz}, {Spitler}, {Hessels}, {Chatterjee},
  {Cordes}, {Kaspi}, {Wharton}, {Bassa}, {Bogdanov}, {Camilo}, {Crawford},
  {Deneva}, {van Leeuwen}, {Lynch}, {Madsen}, {McLaughlin}, {Mickaliger},
  {Parent}, {Patel}, {Ransom}, {Seymour}, {Stairs}, {Stappers}, \&
  {Tendulkar}}]{ssh+16b}
{Scholz}, P., {Spitler}, L.~G., {Hessels}, J.~W.~T., {et~al.} 2016, \apj, 833,
  177

\bibitem[{{Scholz} {et~al.}(2017){Scholz}, {Bogdanov}, {Hessels}, {Lynch},
  {Spitler}, {Bassa}, {Bower}, {Burke-Spolaor}, {Butler}, {Chatterjee},
  {Cordes}, {Gourdji}, {Kaspi}, {Law}, {Marcote}, {McLaughlin}, {Michilli},
  {Paragi}, {Ransom}, {Seymour}, {Tendulkar}, \& {Wharton}}]{sbh+17}
{Scholz}, P., {Bogdanov}, S., {Hessels}, J.~W.~T., {et~al.} 2017, ArXiv
  e-prints, arXiv:1705.07824

\bibitem[{{Shannon} {et~al.}(2018){Shannon}, {Macquart}, {Bannister}, {Ekers},
  {James}, {Os{\l}owski}, {Qiu}, {Sammons}, {Hotan}, {Voronkov}, {Beresford},
  {Brothers}, {Brown}, {Bunton}, {Chippendale}, {Haskins}, {Leach},
  {Marquarding}, {McConnell}, {Pilawa}, {Sadler}, {Troup}, {Tuthill},
  {Whiting}, {Allison}, {Anderson}, {Bell}, {Collier}, {G{\"u}rkan}, {Heald},
  \& {Riseley}}]{smb+18}
{Shannon}, R.~M., {Macquart}, J.-P., {Bannister}, K.~W., {et~al.} 2018, \nat,
  562, 386

\bibitem[{{Spitler} {et~al.}(2014){Spitler}, {Cordes}, {Hessels}, {Lorimer},
  {McLaughlin}, {Chatterjee}, {Crawford}, {Deneva}, {Kaspi}, {Wharton},
  {Allen}, {Bogdanov}, {Brazier}, {Camilo}, {Freire}, {Jenet},
  {Karako-Argaman}, {Knispel}, {Lazarus}, {Lee}, {van Leeuwen}, {Lynch},
  {Ransom}, {Scholz}, {Siemens}, {Stairs}, {Stovall}, {Swiggum},
  {Venkataraman}, {Zhu}, {Aulbert}, \& {Fehrmann}}]{sch+14}
{Spitler}, L.~G., {Cordes}, J.~M., {Hessels}, J.~W.~T., {et~al.} 2014, \apj,
  790, 101

\bibitem[{{Spitler} {et~al.}(2016){Spitler}, {Scholz}, {Hessels}, {Bogdanov},
  {Brazier}, {Camilo}, {Chatterjee}, {Cordes}, {Crawford}, {Deneva}, {Ferdman},
  {Freire}, {Kaspi}, {Lazarus}, {Lynch}, {Madsen}, {McLaughlin}, {Patel},
  {Ransom}, {Seymour}, {Stairs}, {Stappers}, {van Leeuwen}, \& {Zhu}}]{ssh+16a}
{Spitler}, L.~G., {Scholz}, P., {Hessels}, J.~W.~T., {et~al.} 2016, \nat, 531,
  202

\bibitem[{{Spitler} {et~al.}(2018){Spitler}, {Herrmann}, {Bower}, {Chatterjee},
  {Cordes}, {Hessels}, {Kramer}, {Michilli}, {Scholz}, {Seymour}, \&
  {Siemion}}]{shb+18}
{Spitler}, L.~G., {Herrmann}, W., {Bower}, G.~C., {et~al.} 2018, ArXiv
  e-prints, arXiv:1807.03722

\bibitem[{{Tendulkar} {et~al.}(2017){Tendulkar}, {Bassa}, {Cordes}, {Bower},
  {Law}, {Chatterjee}, {Adams}, {Bogdanov}, {Burke-Spolaor}, {Butler},
  {Demorest}, {Hessels}, {Kaspi}, {Lazio}, {Maddox}, {Marcote}, {McLaughlin},
  {Paragi}, {Ransom}, {Scholz}, {Seymour}, {Spitler}, {van Langevelde}, \&
  {Wharton}}]{tbc+17}
{Tendulkar}, S.~P., {Bassa}, C.~G., {Cordes}, J.~M., {et~al.} 2017, \apjl, 834,
  L7

\bibitem[{{Thornton} {et~al.}(2013){Thornton}, {Stappers}, {Bailes},
  {Barsdell}, {Bates}, {Bhat}, {Burgay}, {Burke-Spolaor}, {Champion}, {Coster},
  {D'Amico}, {Jameson}, {Johnston}, {Keith}, {Kramer}, {Levin}, {Milia}, {Ng},
  {Possenti}, \& {van Straten}}]{tsb+13}
{Thornton}, D., {Stappers}, B., {Bailes}, M., {et~al.} 2013, Science, 341, 53

\bibitem[{{Treumann}(2006)}]{tre06}
{Treumann}, R.~A. 2006, \aapr, 13, 229

\bibitem[{{van Straten} \& {Bailes}(2011)}]{Str11}
{van Straten}, W., \& {Bailes}, M. 2011, \pasa, 28, 1

\bibitem[{{van Straten} {et~al.}(2012){van Straten}, {Demorest}, \&
  {Oslowski}}]{Str12}
{van Straten}, W., {Demorest}, P., \& {Oslowski}, S. 2012, Astronomical
  Research and Technology, 9, 237

\bibitem[{{Watson} \& {Melrose}(2006)}]{wm06}
{Watson}, P.~G., \& {Melrose}, D.~B. 2006, \apj, 647, 1142

\bibitem[{{Waxman}(2017)}]{wax17}
{Waxman}, E. 2017, \apj, 842, 34

\bibitem[{{Williams} \& {Berger}(2016)}]{wb16}
{Williams}, P.~K.~G., \& {Berger}, E. 2016, \apjl, 821, L22

\bibitem[{{Yao} {et~al.}(2017){Yao}, {Manchester}, \& {Wang}}]{ymw17}
{Yao}, J.~M., {Manchester}, R.~N., \& {Wang}, N. 2017, \apj, 835, 29

\bibitem[{{Zarka}(1992)}]{Zar92}
{Zarka}, P. 1992, Advances in Space Research, 12, 99

\bibitem[{{Zhang} {et~al.}(2018){Zhang}, {Gajjar}, {Foster}, {Siemion},
  {Cordes}, {Law}, \& {Wang}}]{zgf+18}
{Zhang}, Y.~G., {Gajjar}, V., {Foster}, G., {et~al.} 2018, \apj, 866, 149

\end{thebibliography}



\newpage


\begin{table*}
    \begin{center}
    \scriptsize
    \caption{\footnotesize Properties of detected bursts.
      Uncertainties are the 68\% confidence interval, unless otherwise stated.
    \label{tab:bursts}}
    \begin{tabular}{llllllll}
    \hline
    \hline
    ID$^a$ & Barycentric         & Peak Flux        & Fluence      & $W_{\rm sb}$ & Drift Rate 	 & DM Max. $({\rm d}l/{\rm d}t)^2$  & DM Peak S/N      \\
        & Peak Time (MJD)$^b$ & Density (Jy)$^c$ & (Jy\,ms)$^c$ & (ms)$^d$        & (MHz\,ms$^{-1}$)$^e$ & (pc\,cm$^{-3}$)$^f$ & (pc\,cm$^{-3}$)$^g$    \\
    \hline
AO-00 & 57364.2046326656 & 0.03	     	 &	0.1     	 &	$\cdots$	& $\cdots$ & $\cdots$ & 557.7(2)	\\
AO-01 & 57638.4659716231 & 0.3	     	 &	0.6		     &	1.03(5)			&	$-204.6$(1)			& $\cdots$  & 561.50(2)	\\
AO-02 & 57638.4675640004 & 0.4	     	 &	0.6		     &	0.19(1)			&	$-122.95$(1)			& 560.68(2) 	& 562.96(2)	\\
AO-03 & 57640.4138405217 & 0.1	     	 &	0.2		     &	0.25(5)			&	$-187.67$(3)			& $\cdots$ & 562.24(2)	\\
AO-04 & 57641.4594528637 & 0.2	     	 &	0.2		     &	0.3(3)			&	$-221.0$(1)		& $\cdots$  & 562.24(2)	\\
AO-05 & 57641.4645632098 & 1.0	     	 &	6.2		     &	0.34(4)			&	$-46.135$(3)			& 560.60(3)		& 565.85(2)	\\
AO-06 & 57642.4715691734 & 0.2	     	 &	0.6		     &	0.31(1)			&	$-129.04$(2)			& 560.50(2)		& 562.66(2)	\\
AO-07 & 57642.4754649610 & 0.4	     	 &	1.1		     &	0.24(2)			&	$-128.81$(1)		& 560.50(3)& 562.83(2)	\\
AO-08 & 57644.4110709268 & 0.2	     	 &	0.3		     &	0.43(3)			&	$-140.76$(1)			& $\cdots$ & 562.16(2)				     \\
AO-09 & 57646.4173141213 & 0.1	     	 &	0.3		     &	0.2(2)			&	$-205.75$(3)			& $\cdots$ & 561.17(5)				     \\
AO-10 & 57646.4278138709 & 0.4	     	 &	0.9		     &	0.23(1)			&	$-50.11$(1)			& 560.50(3) & 562.52(2)				     \\
AO-11 & 57648.4307890113 & 0.3	     	 &	0.6		     &	0.14(2)			&	$\sim$0			& 560.55(3) & 560.74(2)				     \\
AO-12 & 57648.4581115606 & 0.2	     	 &	0.2		     &	0.35(2)			&	$-168.55$(8)			& 560.53(3) & 561.68(2)				     \\
AO-13 & 57649.4281585259 & 0.2	     	 &	0.6		     &	0.17(3)			&	$-286.89$(3)		&  560.67(4) & 561.38(2)			     \\
GB-01 & 57647.2964919448 & 0.4	     	 &	0.5		     &	0.13(1)			& 	$-237.40$(4)				& 560.79(1)&	564.21(4)			     \\
GB-02 & 57649.3337214719 & 0.2	     	 &	0.4		     &	0.16(1)			&	$-251.70$(3)			& 560.65(1) &	563.96(4)			     \\
GB-03 & 57927.5700691158 & 0.05	     	 &	0.1	     	 &	0.30(1)			&	$-141.9$(1)		& 560.5(1)&	567.27(8)			     \\
GB-04 & 57928.7263586936 & 0.05	     	 &	0.1	     	 &	0.4(1)			&	$-276.3$(1)			& 560.1(1) &	563.10(7)			     \\
GB-BL & 57991.5765740056 & 0.4	     	 &	0.5		     &	0.13(1)			& 	$-865.5$(2)				& 563.86(5)&	595.1(4)			     \\
	\hline
    \hline
    \newline
    \end{tabular}
    \end{center}
    \begin{footnotesize}
	$^a$ Central observing frequencies: AO-00 to AO-13: 1.4\,GHz; GB-01 to GB-04: 2.0\,GHz; GB-BL: 6.5\,GHz.\\
	$^b$ Arrival time of the centroid of the full-burst envelope, corrected to the Solar System Barycenter and referenced to infinite frequency (i.e the time delay due to dispersion is removed) using an assumed DM$ = 560.5$\,pc\,cm$^{-3}$. \\
    $^c$ Given for the brightest sub-burst.\\
    $^d$ The characteristic sub-burst durations determined from the ACF analysis.\\
    $^e$ Best-fit linear trend to the sub-burst centroids.  A negative sign is used to indicate decreasing frequency.\\
	$^f$ DM at which the squared time-derivative or the profile is maximized.\\
	$^g$ DM at which the peak S/N is maximized.\\
    \end{footnotesize}
\end{table*}

\newpage


\newpage


\newpage


\newpage


\newpage


\end{document}